\documentclass[review,12pt]{elsarticle}

\usepackage{amssymb}
\usepackage{amsthm}
\usepackage{framed} 
\usepackage{amsmath,color}
\usepackage{mathrsfs}
\usepackage{graphicx}
\usepackage{epstopdf}
\usepackage{float}
\usepackage{caption}
\usepackage{subcaption}
\usepackage{bm}
\usepackage{bbm}
\usepackage{mathrsfs}
\usepackage{hyperref}
\usepackage{cleveref}
\usepackage{algorithm}
\usepackage{algpseudocode}
\usepackage{enumitem}

\usepackage{soul}
\usepackage{accents}
\usepackage{color,soul} 
\usepackage{color} 
\usepackage{bm}
\usepackage{bbm}
\usepackage{upgreek} 
\usepackage{cprotect} 
\usepackage[margin=2cm]{geometry}
\biboptions{sort&compress}
\soulregister\citep7 
\soulregister\citet7 
\soulregister\citealp7 
\newcommand{\refe}[1]{(\ref{#1})}
\newsavebox{\measurebox} 
\usepackage{titlesec} 
\usepackage[T1]{fontenc} 
\usepackage{lmodern} 
\input{glyphtounicode}  
\newcommand{\tm}{\textrm}
\def\onedot{$\mathsurround0pt\ldotp$}
\def\cddot{
  \mathbin{\vcenter{\baselineskip.67ex
    \hbox{\onedot}\hbox{\onedot}}%
  }}%
\journal{European Journal of Mechanics A/Solids}

\makeatletter
\def\@author#1{\g@addto@macro\elsauthors{\normalsize%
    \def\baselinestretch{1}%
    \upshape\authorsep#1\unskip\textsuperscript{%
      \ifx\@fnmark\@empty\else\unskip\sep\@fnmark\let\sep=,\fi
      \ifx\@corref\@empty\else\unskip\sep\@corref\let\sep=,\fi
      }%
    \def\authorsep{\unskip,\space}%
    \global\let\@fnmark\@empty
    \global\let\@corref\@empty  
    \global\let\sep\@empty}%
    \@eadauthor={#1}
}
\makeatother

\titleformat{\paragraph}
{\normalfont\normalsize\itshape}{\theparagraph}{1em}{}
\titlespacing*{\paragraph}
{0pt}{3.25ex plus 1ex minus .2ex}{1.5ex plus .2ex}

\begin{document}

\begin{frontmatter}



\title{Accelerated high-cycle phase field fatigue predictions}


\author{Philip K. Kristensen \fnref{DTU}}

\author{Alireza Golahmar \fnref{DTU,ICL}}

\author{Emilio Mart\'{\i}nez-Pa\~neda\fnref{ICL}}

\author{Christian F. Niordson\corref{cor1}\fnref{DTU}}
\ead{cfni@dtu.dk}

\address[DTU]{Department of Civil and Mechanical Engineering, Technical University of Denmark, DK-2800 Kgs. Lyngby, Denmark}

\address[ICL]{Department of Civil and Environmental Engineering, Imperial College London, London SW7 2AZ, UK}

\cortext[cor1]{Corresponding author.}

\begin{abstract}
Phase field fracture models have seen widespread application in the last decade. Among these applications, its use to model the evolution of fatigue cracks has attracted particular interest, as fatigue damage behaviour can be predicted for arbitrary loading histories, dimensions and complexity of the cracking phenomena at play. However, while cycle-by-cycle calculations are remarkably flexible, they are also computationally expensive, hindering the applicability of phase field fatigue models for technologically-relevant problems. In this work, a computational framework for accelerating phase field fatigue calculations is presented. Two novel acceleration strategies are proposed, which can be used in tandem and together with other existing acceleration schemes from the literature. The computational performance of the proposed methods is documented through a series of 2D and 3D boundary value problems, highlighting the robustness and efficiency of the framework even in complex fatigue problems. The observed reduction in computation time using both of the proposed methods in tandem is shown to reach a speed-up factor of 32, with a scaling trend enabling even greater reductions in problems with more load cycles.\\ 
\end{abstract}


\end{frontmatter}


\section{Introduction}\label{sec:Introduction}
The phase field fracture model has received substantial attention in the last decade - and for good reason. The original model, proposed by Bourdin \textit{et al.} \cite{Bourdin2000} as a regularization of the variational fracture formulation by Francfort and Marigo \cite{Francfort1998}, is flexible and simple to implement numerically. It can readily capture complex cracking phenomena such as crack branching \cite{Borden2012}, coalescence \cite{Kristensen2020b}, complex crack trajectories \cite{Hirshikesh2019} and crack nucleation from non-sharp defects \cite{Tanne2018}. Moreover, it can naturally capture the crack size effect \cite{Tanne2018,Kristensen2021} and be readily extended to accommodate specific failure surfaces \cite{Navidtehrani2022,DeLorenzis2022}. The model has proven to be extremely versatile and thus has been used in a vast number of applications, both within complex fracture problems such as cohesive fracture \cite{Wu2017,Feng2022}, micromechanical damage \cite{Guillen-Hernandez2020,Tan2021}, and ductile fracture \cite{Aldakheel2018, Alessi2018}, but also in multi-physics applications ranging from thermal shocks \cite{Bourdin2014} and moisture effects \cite{Ye2022} to hydrogen embrittlement \cite{Martinez-Paneda2018,Duda2018,Anand2019,Kristensen2020} and Lithium-ion battery degradation \cite{Klinsmann2016,Boyce2022,Ai2022}.\\

Among the many problems attracting phase field developments, fatigue is arguably one of the most important ones, from both scientific and technological perspectives. Using phase field as a framework for fatigue models is an attractive prospect \cite{Alessi2023}, as fatigue remains a longstanding challenge in solid mechanics and a sufficiently flexible phase field fatigue model could readily encompass the aforementioned phase field models and applications. An example of such a flexible framework is found in the work by Carrara \textit{et al.} \cite{Carrara2020}, where a history variable is introduced to introduce a dependence of the fracture energy on the loading history of the material. The model naturally recovers the Paris Law and S-N curve behaviour and has proven an attractive platform from which to develop phase field fatigue models for various applications \cite{Loew2020,Golahmar2022,Simoes2021,Simoes2022}. Aside from Ref. \cite{Carrara2020}, there have been several other notable works, both of a cycle-by-cycle nature \cite{Seiler2019,Mesgarnejad2019,Song2022} and more practical approaches, which take the Paris behaviour as an input \cite{Lo2019}. \\

Two known drawbacks associated with phase field fracture are the computational cost associated with the need for a sufficiently fine mesh to resolve the phase field length scale \cite{Kristensen2021}, and the inefficiency of the solution due to the non-convexity of the balance equations \cite{Gerasimov2016}. Significant efforts have been extended towards remedying both issues. Strategies to ease meshing requirements include adaptive mesh refinement \cite{Heister2015,Klinsmann2015, Freddi2022, Freddi2023}, specialised element formulations \cite{Olesch2021} and the use of a combined finite element-finite volume approach \cite{Sargado2021}. Similarly, a wealth of improvements have also been proposed for the solution strategy, including (residual) control algorithms \cite{Ambati2016,Seles2019}, quasi-Newton methods \cite{Kristensen2020a,Wu2019}, line search algorithms \cite{Gerasimov2016,Lampron2021,Borjesson2022}, and multigrid approaches \cite{JodlBauer2020}. The computational costs of phase field fracture become particularly demanding when performing cycle-by-cycle computations of fatigue, hindering high cycle fatigue analyses. A common strategy to minimize the costs of computing high cycle fatigue is the use of so-called \textit{cycle jumping}, where the cycle-by-cycle solution is extrapolated to skip the computation of several load cycles \cite{Cojocaru2006}. Loew \textit{et al.} \cite{Loew2020} introduced such a scheme for phase field fatigue by locally extrapolating the fatigue history variable.  \\

This paper seeks to introduce alternative means of accelerating cycle-by-cycle phase field fatigue computations. Two methods are proposed which are mutually compatible and individually provide substantial computational performance improvements. Furthermore, neither methods prevent the use of existing cycle jump strategies, which can be included for additional performance improvements. The first computational acceleration method proposed is a modification to existing staggered solution strategies commonly adopted in the phase field literature \cite{Miehe2010,Ambati2014,Seles2019}, so that the tangent stiffness matrices are not updated in each load step, but rather stored in factorized form such that subsequent increments and iterations are solved with a significant reduced computational cost. This approach, henceforth referred to as \emph{Modified Newton (MN)}, is especially suitable for high-cycle fatigue, where very small changes to the overall system are observed between individual load increments. The second method proposed, referred to as \emph{Constant Load Accumulation (CLA)}, is suitable for problems where only one step of the loading cycle contributes significantly to the fatigue accumulation. For systems where this assumption is valid, the fatigue accumulation rule can be adjusted to permit the simplification of the loading curve to a single load increment per cycle, significantly reducing the total number of increments in the simulation. As shall be shown, for high-cycle problems this approach can be extended to capture multiple load cycles in a single increment with negligible loss of accuracy for even greater computational
performance improvement. \\

The manuscript is organized as follows. Section \ref{sec:Theory} formulates the phase field fatigue framework used for the cycle-by-cycle computations. The details of the numerical aspects of the finite element solution and the proposed methods are given in Section \ref{sec:FE}. Subsequently, a series of numerical examples are analyzed in Section \ref{sec:Results} to illustrate the capabilities of the proposed method. Concluding remarks end the paper in Section \ref{sec:Conclusions}.

\section{A phase field model for fatigue}
\label{sec:Theory}

This section introduces the specific phase field fatigue model adopted, which is based on the work by Carrara \textit{et al.} \cite{Carrara2020}. This choice is grounded on its flexibility, simplicity of implementation and suitability to be used in conjunction with other acceleration schemes (such as those of Ref. \cite{Loew2020}). Furthermore, the fatigue acceleration strategies presented here can be readily incorporated into a wide range of phase field fatigue models and, as such, the specific choice phase field model is of secondary importance. 

\subsection{Basic theory}
Consider a solid domain $\Omega \in \mathbb{R}^n$ with boundary $\partial\Omega \in \mathbb{R}^{n-1}$. In a small deformations context, we consider a displacement field $\mathbf{u}\in \mathbb{R}$ and a phase field $\phi\in [0;1]$, for which a value of $0$ denotes intact material and a value of $1$ denotes broken material with vanishing stiffness. Then, the standard so-called \verb|AT2| phase field fracture model \cite{Bourdin2000} can be formulated from the minimisation of the following energy functional: 
\begin{equation}
\label{eq:AT2Energy}
    \mathcal{E}(\bm{\varepsilon}(\mathbf{u}),\phi,\nabla\phi) = \int_\Omega \left[\psi(\bm{\varepsilon}(\mathbf{u}),\phi)+\frac{G_c}{2}\left(\dfrac{\phi^2}{\ell} + \ell \nabla\phi\cdot\nabla\phi\right)\right] \,\tm{d}V
\end{equation}
where $G_c$ is the critical energy release rate or material toughness, $\ell$ denotes the phase field length scale, and $\psi(\bm{\varepsilon}(\mathbf{u}),\phi)$ is the strain energy density, which for a linear elastic solid may be expressed as; 
\begin{equation}
    \psi(\bm{\varepsilon}(\mathbf{u}),\phi)= (1-\phi)^2\psi_0(\mathbf{u}) = (1-\phi)^2\frac{1}{2}\bm{\varepsilon}\cddot\mathbf{C}_0\cddot\bm{\varepsilon}.
\end{equation}
Here, $\mathbf{C}_0$ is the linear elastic stiffness tensor of the material and $\bm{\varepsilon}$ is the infinitesimal strain tensor, given by $\bm{\varepsilon}=(\nabla\mathbf{u}+\nabla\mathbf{u}^T)/2$. The phase field variable, $\phi$, is seen to degrade the material stiffness.

\subsection{Extension to fatigue}
To extend the above fracture framework to account for fatigue damage, Carrara and co-workers \cite{Carrara2020} proposed introducing a degradation function $f(\Bar{\alpha})$, which reduces the material toughness, as a function of an accumulated fatigue history variable $\Bar{\alpha}$. Several options are available for both the formulation of the degradation function and the accumulated fatigue history variable. Here, as in Ref. \cite{Seles2021}, the accumulated fatigue history variable at time step $n+1$ is introduced as the cumulative positive increments of undegraded elastic strain energy density: 
\begin{equation}\label{eq:Faccumulation}
    \Bar{\alpha}_{n+1}=\Bar{\alpha}_n+|\psi_{0,n+1}-\psi_{0,n}|H(\psi_{0,n+1}-\psi_{0,n}),
\end{equation}
where $H$ is the Heaviside function. Also following Ref. \cite{Carrara2020}, we use the asymptotic fatigue degradation function:
\begin{equation}
    f(\Bar{\alpha})=\begin{cases} \hspace{0.8cm}0 &\tm{if   } \Bar{\alpha}\leq \alpha_T \\
    \left(\dfrac{2\alpha_T}{\Bar{\alpha}+\alpha_T}\right)^2 &\tm{else   }
    \end{cases}.
\end{equation}
\noindent where the threshold parameter $\alpha_T$ introduces a lower limit below which accumulated fatigue does not influence the material toughness. This threshold is here chosen as $\alpha_T=G_c/12\ell$.

\subsection{Principle of Virtual Power}
Let us first define the Cauchy stress tensor $\bm{\sigma}$ in terms of the strain energy density $\psi$,
\begin{equation}
    \bm{\sigma}= \frac{\partial \psi}{\partial \bm{\varepsilon}} = (1-\phi)^2\bm{\sigma}_0 =  (1-\phi)^2\mathbf{C}_0\cddot\bm{\varepsilon}.
\end{equation}
Then, the internal energy of the system can be expressed as
\begin{equation}
    \mathcal{W}= \int_\Omega (1-\phi)^2\bm{\sigma}_0\cddot\bm{\varepsilon}\,\tm{d}V + \int_\Omega\int_0^t f(\Bar{\alpha}(\tau))\dfrac{G_c}{\ell}\left(\phi\Dot{\phi}+\ell^2\nabla\phi\cdot\nabla\Dot{\phi}\right)\, \tm{d}\tau\,\tm{d}V.
\end{equation}
Alternatively, the above may be expressed in terms of internal power density as
\begin{equation}
    \Dot{\mathcal{W}} = \int_\Omega\left[(1-\phi)^2\bm{\sigma}_0\cddot\Dot{\bm{\varepsilon}}+\frac{\partial \psi}{\partial\phi} \Dot{\phi}+ f(\Bar{\alpha}(t))\dfrac{G_c}{\ell}\left(\phi\Dot{\phi}+\ell^2\nabla\phi\cdot\nabla\Dot{\phi}\right)\right]\,\tm{d}V.
\end{equation}
The external power depends only on the external mechanical loading 
\begin{equation}
 \Dot{\mathcal{F}}= \int_\Omega   \mathbf{b}\cdot\Dot{\mathbf{u}}\, \tm{d}V + \int_{\partial\Omega_t}\mathbf{t}\cdot\Dot{\mathbf{u}}\,\tm{d}S,
\end{equation}
where $\partial\Omega_t$ denotes the part of the boundary where mechanical tractions $\mathbf{t}$ are applied and $\mathbf{b}$ are the body forces. Here, the external contributions due to the phase field variable and its work-conjugate are omitted, as only problems with homogeneous phase field boundary conditions are considered. The balance of virtual power requires
\begin{equation}
    \Dot{\mathcal{W}}- \Dot{\mathcal{F}}=0,
\end{equation}
which, after applying integration by parts, may be expressed as: 
\begin{equation}
\begin{split}
    &\int_\Omega -\left[(1-\phi)^2\nabla \cdot \bm{\sigma}_0+\mathbf{b}\right]\cdot\Dot{\mathbf{u}}\, \tm{d}V +\int_\Omega \left\{ \frac{\partial\psi}{\partial\phi}-\frac{G_c}{\ell}\left[f(\Bar{\alpha})\left(\ell^2\nabla^2 \phi  -\phi\right)  + \ell^2\nabla f(\Bar{\alpha}) \cdot\nabla \phi\right] \right\} \Dot{\phi} \, \tm{d}V\\
    &+G_c\ell\int_{\partial\Omega}f(\Bar{\alpha})\nabla \phi\cdot\mathbf{n}\Dot{\phi}\, \tm{d}S+\int_{\partial\Omega_t}\left[(1-\phi)^2\bm{\sigma}\cdot \mathbf{n}-\mathbf{t}\right]\cdot \Dot{\mathbf{u}}\, \tm{d}S =0
    \end{split}
\end{equation}
The above must hold for arbitrary, kinematically admissible, variations of the velocities $\Dot{\mathbf{u}}$ and phase field increments $\Dot{\phi}$, which by standard arguments implies the following local balance equations and accompanying boundary conditions: 
\begin{align}\label{eq:StrongFormDisp}
    \nabla \cdot \left[(1-\phi)^2 \bm{\sigma}_0 \right] +\mathbf{b} = 0\hspace{9.0cm} &\tm{in }\Omega\\ \label{eq:StrongFormPhi}
    -2(1-\phi)\psi_0+f(\Bar{\alpha})\dfrac{G_c}{\ell}\left( \phi-\ell^2\nabla^2 \phi\right)-G_c\ell\nabla f(\Bar{\alpha})\cdot\nabla\phi = 0\hspace{2.6cm} &\tm{in }\Omega\\ \label{eq:StrongBCDisp}
    (1-\phi)^2\bm{\sigma}_0\cdot \mathbf{n}= \mathbf{t}\hspace{9.9cm}   &\tm{on }\partial\Omega\\ \label{eq:StrongBCPhi}
    \nabla\phi\cdot \mathbf{n}=0\hspace{11.1cm} &\tm{on }\partial\Omega
\end{align}

\subsection{Strain energy split to adequately handle compression behaviour}

In its original formulation, the phase field fracture model predicts a symmetric behaviour under tension and compression. That is, crack growth is equally driven by compressive and tensile stresses and, since the degradation of the material stiffness is similarly isotropic, the crack faces are allowed to interpenetrate and while carrying no compressive loads. A common strategy to mitigate this is to decompose the strain energy density into active and passive parts such that: 
\begin{equation}
    \psi = (1-\phi)^2 \psi_0^+(\mathbf{u}) + \psi_0^-(\mathbf{u})
\end{equation}
where only the active part of the strain energy density ($\psi_0^+$) contributes to crack growth and only the active part of the stiffness is degraded by the phase field variable. Several suggestions have been made for defining the active and passive parts of the strain energy density, with the two most popular being the volumetric/deviatoric split by Amor \textit{et al.} \cite{Amor2009} and the spectral split by Miehe \textit{et al.}  \cite{Miehe2010a}. The volumetric/deviatoric split is given by
\begin{equation}
    \begin{split}
        &\psi_0^+ = \frac{1}{2}K\left\langle \tm{tr}\bm\varepsilon\right\rangle_+^2 + \mu (\bm{\varepsilon}_{dev}\cddot \bm{\varepsilon}_{dev}) \\
        &\psi_0^- = \frac{1}{2}K\left\langle \tm{tr}\bm\varepsilon\right\rangle_-^2,
    \end{split}
\end{equation}
where $\left\langle{~}\right\rangle_\pm$ denotes the two signed Macaulay brackets, $\bm{\varepsilon}_{dev}$ is the deviatoric part of the strain tensor, $K$ is the bulk modulus and $\mu$ is the shear modulus or second Lamé parameter. On the other side, the spectral split is based on a spectral decomposition of the strain tensor: $\bm{\varepsilon}_\pm=\sum_{a=1}^3 \left\langle \varepsilon_I\right\rangle_{\pm} \mathbf{n}_I \otimes \mathbf{n}_I$, where $\varepsilon_I$ are the principal strains and $\mathbf{n}_I$ denote the principal strain  directions (with $I=1,2,3$). The, the spectral strain energy decomposition is defined as
\begin{equation}
    \psi^\pm_0 = \frac{1}{2}\lambda\left\langle \tm{tr}\bm{\varepsilon}\right\rangle^2_{\pm} + \mu\tm{tr} \left(\bm{\varepsilon}^2_\pm \right),
\end{equation}
with $\lambda$ denoting the first Lamé parameter and $\tm{tr}$ being the trace operator.\\ 

A significant improvement to the numerical performance of these splits was introduced with the so-called hybrid scheme by Ambati and co-workers \cite{Ambati2014}, where only the active part of the strain energy contributes to crack growth, but the stiffness is isotropically degraded by damage, with the caveat that degradation only applies if the stress state is predominantly tensile. An alternative strain energy decomposition, which has been shown to be particularly effective for fatigue modelling \cite{Golahmar2023}, is the so-called no-tension split by Freddi and Royer-Carfagni \cite{Freddi2010}. The no-tension split, first intended for masonry-like materials, filters out contributions from compressive strains more effectively than other approaches. Using $\lambda$ and $\mu$ to denote the Lamé parameters, $E$ and $\nu$ respectively being Young's modulus and Poisson's ratio, and taking $\varepsilon_1, \, \varepsilon_2, \, \varepsilon_3$ as the principal strains, with $\varepsilon_1$ being the largest, the strain energy decomposition is given as \cite{Lo2019}

\begin{normalsize}
\begin{equation}\label{eq:Freddi}
\begin{alignedat}{4}
&\tm{if} \hspace{1cm}  &&\varepsilon_3 >0 &&\tm{then} &&\begin{cases}
        &\psi^+_0 = \frac{\lambda}{2}\left(\varepsilon_1+\varepsilon_2+\varepsilon_3 \right)^2+ \mu \left(\varepsilon_1^2+\varepsilon_2^2 + \varepsilon_3^2 \right) \\
        &\psi^-_0 = 0
    \end{cases} \\
&\tm{elseif} &&\varepsilon_2+\nu\varepsilon_3 > 0 &&\tm{then } &&\begin{cases}
        &\psi^+_0 = \frac{\lambda}{2}\left(\varepsilon_1+\varepsilon_2+2\nu \varepsilon_3 \right)^2+ \mu \left[\left(\varepsilon_1+\nu\varepsilon_3 \right)^2+\left(\varepsilon_2+\nu\varepsilon_3\right)^2\right] \\
        &\psi^-_0 = \frac{E}{2}\varepsilon_3^2
    \end{cases} \\
&\tm{elseif} &&(1- \nu) \varepsilon_1+\nu(\varepsilon_2 + \varepsilon_3) > 0 \hspace{0.5cm}  &&\tm{then  } &&\begin{cases}
        &\psi^+_0 = \frac{\lambda}{2}\left[\left(1-\nu\right)\varepsilon_1+\nu\varepsilon_2+\nu \varepsilon_3 \right]^2 \\
        &\psi^-_0 = \frac{E}{2\left((1-\nu^2\right)}\left(\varepsilon_2^2+\varepsilon^2_3+2\nu\varepsilon_2\varepsilon_3\right)
    \end{cases} \\
    & && &&\tm{else } &&\begin{cases}
        &\psi^+_0 = 0 \\
        &\psi^-_0 = \frac{\lambda}{2}\left(\varepsilon_1+\varepsilon_2+\varepsilon_3 \right)^2+ \mu \left(\varepsilon_1^2+\varepsilon_2^2 + \varepsilon_3^2 \right) 
    \end{cases}
\end{alignedat}
\end{equation}
\end{normalsize}

Unless otherwise stated, this no-tension split by Freddi and Royer-Carfagni \cite{Freddi2010} is the one adopted in the numerical experiments reported in this manuscript. 

\section{Finite element implementation}
\label{sec:FE}

This section provides details of the numerical implementation of the phase field fatigue model presented in Section \ref{sec:Theory}. The finite element method is used and the solution of the resulting system of equations is discussed, together with the fatigue acceleration methods presented in this work: the Modified Newton (MN) and the Constant Load Accumulation (CLA) solution strategies. The implementation is carried out using the Ferrite.jl finite element library \cite{Carlsson_Ferrite_jl_2021} \footnote{The Julia implementation developed is openly shared with the community and made available to download at \url{www.empaneda.com/codes}.}. 

\subsection{Crack irreversibility}

Enforcing damage irreversibility is of critical importance when considering non-monotonic loading. For simplicity, we shall here make use of the so-called history field $\mathcal{H}$ approach pioneered by Miehe et al. \cite{Miehe2010}. Accordingly, the history field is defined as the maximum active strain energy density experienced in a point during the loading history
\begin{equation}
    \mathcal{H} =\max\limits_{\tau \in[0,t]}\psi^+_0( \tau)
\end{equation}
and it replaces the active undegraded strain energy density $\psi_0^+$ as the crack driving force in the phase field equation (\ref{eq:StrongFormPhi}). While this approach is convenient and tends to ease the convergence of the phase field equations, it has also been the target of sensible objections \cite{Linse2017,Strobl2020}, especially regarding its influence on crack nucleation from non-sharp defects and its non-variational nature. The latter issue is of little relevance here, as the fatigue extension of phase field is not variationally consistent in the form adopted here. A more critical aspect in the case of fatigue is that for variable amplitude loading, only the locally maximal loads will be retained as a crack driving force throughout cycles that also include lower loads. However, this scenario is not relevant to this work, as the numerical examples deal with constant amplitude loading and pre-existing sharp defects.\\

Another method of enforcing irreversibility of fully formed cracks is the so-called crack-set method by Bourdin \textit{et al.} \cite{Bourdin2000}, where nodes in which the phase field exceeds a given threshold are added to a set of nodes subject to a $\phi = 1$ Dirichlet condition. Anecdotally, we find that this method seems to ease some convergence issues which have been observed to occur at the original crack tip after some degree of crack growth in high-cycle fatigue simulations, regardless of whether the degraded or undegraded strain energy is used to obtain the fatigue variable. As a result, this work uses both the history variable approach and the crack set method in tandem, with the threshold value for nodes to be added to the crack set chosen as 0.95.


\subsection{Solution strategy}
The governing equations \refe{eq:StrongFormDisp}-\refe{eq:StrongBCPhi} can be reformulated in a numerically convenient decoupled form as
\begin{equation} \label{eq:WeakForm}
    \begin{split}
        \int_{\Omega} \left[ (1-\phi)^2\bm{\sigma}_0 \cddot \delta \bm{\varepsilon} - \mathbf{b} \cdot \delta \mathbf{u} \right] \, \mathrm{d}V + \int_{\partial \Omega_{t}} \mathbf{t} \cdot \delta \mathbf{u}\, \mathrm{d}A &= 0\\[3mm]
        \int_{\Omega} \left[ -2(1-\phi) \psi_0^+\delta \phi +
        f(\Bar{\alpha})G_{c}\left(\dfrac{\phi}{\ell} \delta \phi
        + \ell\nabla \phi \cdot \nabla \delta \phi \right) \right]  \, \mathrm{d}V &= 0
    \end{split}
    \centering
\end{equation}

The weak form equations \refe{eq:WeakForm} are then discretized using standard bilinear elements to form the system of equations: 
\begin{equation}
    {\begin{bmatrix}
        \mathbf{K}^{\mathbf{u}\mathbf{u}} & \mathbf{K}^{\mathbf{u}\mathbf{\phi}}\\[-0.1cm]
       \mathbf{K}^{\mathbf{\phi}\mathbf{u}} & \mathbf{K}^{\phi\phi}
    \end{bmatrix}}{\begin{Bmatrix}
        \textbf{u}\\[-0.1cm] \bm{\phi}
    \end{Bmatrix}}= {\begin{Bmatrix}
        \textbf{r}^{\textbf{u}}\\[-0.1cm] \textbf{r}^{\phi}
    \end{Bmatrix}} 
\end{equation}
where $\mathbf{K}$ and $\mathbf{r}$ are stiffness matrices and residuals vectors, respectively. The phase field equations are non-convex with respect to the variables $\mathbf{u}$ and $\phi$ simultaneously. As a result, the full coupled system is notoriously difficult to solve in a stable and efficient manner (unless unconventional schemes, such as quasi-Newton methods, are used \cite{Kristensen2020a,Wu2019}). However, the equations are convex with respect to the primary variables individually. Therefore, a common strategy is to solve the system in a decoupled fashion, using alternate minimisation. In the following, common staggered schemes are briefly introduced, followed by the proposed modified Newton approach for accelerated fatigue computations. 

\subsubsection{Standard alternate minimization techniques}

Solving the phase field equations by a sequence of alternate minimisation of the two decoupled subproblems was popularized by Miehe et al. \cite{Miehe2010}. The scheme which was then proposed involves solving each of the subproblems independently until individual convergence is achieved before moving on to the next load increment. This is now commonly referred to as a single-pass scheme and introduces significant sensitivity to the size of the load increments. As was shown in Ref. \cite{Kristensen2020a}, this approach may be highly inefficient for fatigue computations. Alternatively, one can use so-called multi-pass schemes where the alternate minimization is repeated until some global convergence criterion is reached. Examples of such convergence criteria can be found in Refs. \cite{Ambati2014, Seles2019}. Here, we adopt the same residual-based multi-pass approach as found in \cite{Lampron2021}, also adopting the tolerances $\texttt{TOL}_\texttt{in} = 10^{-5}$ and $\texttt{TOL}_\texttt{out}=10^{-4}$. The scheme is provided in Algorithm \ref{alg:Multipass} and only differs from the single-pass algorithm by the presence of the \texttt{while} loop.
\begin{algorithm}
\caption{Mutli-pass alternate minimization}
\label{alg:Multipass}
\begin{algorithmic}
\State Increment $n+1$
\State \textbf{Initialize:} $\phi_0 = \phi_n$, $\mathbf{u}_0 = \mathbf{u}_n$, $k=0$

\While{$||R_\phi(\mathbf{u}_{k+1}, \phi_{k+1})||_{\infty} \leq \texttt{TOL}_\texttt{out}$}
\State\textbf{Find} $\phi_{k+1}$ such that $||R_\phi\left(\mathbf{u}_k,\phi_{k+1}\right)||_\infty \leq \texttt{TOL}_\texttt{in}$
\State\textbf{Find} $\mathbf{u}_{k+1}$ such that $||R_\mathbf{u}\left(\mathbf{u}_{k+1},\phi_{k+1}\right)||_\infty \leq \texttt{TOL}_\texttt{in}$
\State $k \gets 1$
\EndWhile
\State $\phi_{n+1} = \phi_k$, $\mathbf{u}_{n+1} = \mathbf{u}_k$
\end{algorithmic}
\end{algorithm}

\subsubsection{Modified Newton approach for accelerated fatigue computations} \label{sec:ModifiedNewton}

In order to accelerate high cycle fatigue computations, a simple modified Newton approach is introduced. In high cycle fatigue computations, it is generally reasonable to expect that changes to the solution variables will be small between individual load increments. Consequently, changes to the tangent stiffness matrices of the system are also expected to be small. As Newton-Raphson based methods do not require the tangent stiffness matrix to be exact, we here propose to modify the multi-pass staggered algorithm given in Algorithm 1, such that the tangent stiffness is only updated and factorized on an as-needed basis. In this implementation, the tangent stiffness matrices for the two subproblems are updated and factorized if any of the following conditions are met:
\begin{itemize}[topsep=8pt,itemsep=0pt,partopsep=1pt, parsep=4pt]
    \item Start of analysis
    \item One of the subproblems fails to converge in $n_i$ inner Newton iterations. 
    \item A number of load increments $n_c$ have passed without updating the stiffness matrices.
\end{itemize}
The parameters $n_i$ and $n_c$ may be chosen differently. While we have not attempted a systematic study of optimal values, which will depend on the size of the boundary value problem and the total number of cycles to failure, the results obtained (see Section \ref{sec:Results}) suggest that choosing $n_c$ to have a higher magnitude in the damage sub-problem will likely result in an improved performance.


\subsection{Accelerating calculations by accumulating fatigue damage under a constant load}
\label{sec:CLA}

\begin{figure}[h]
    \centering
    \includegraphics[width=1\linewidth]{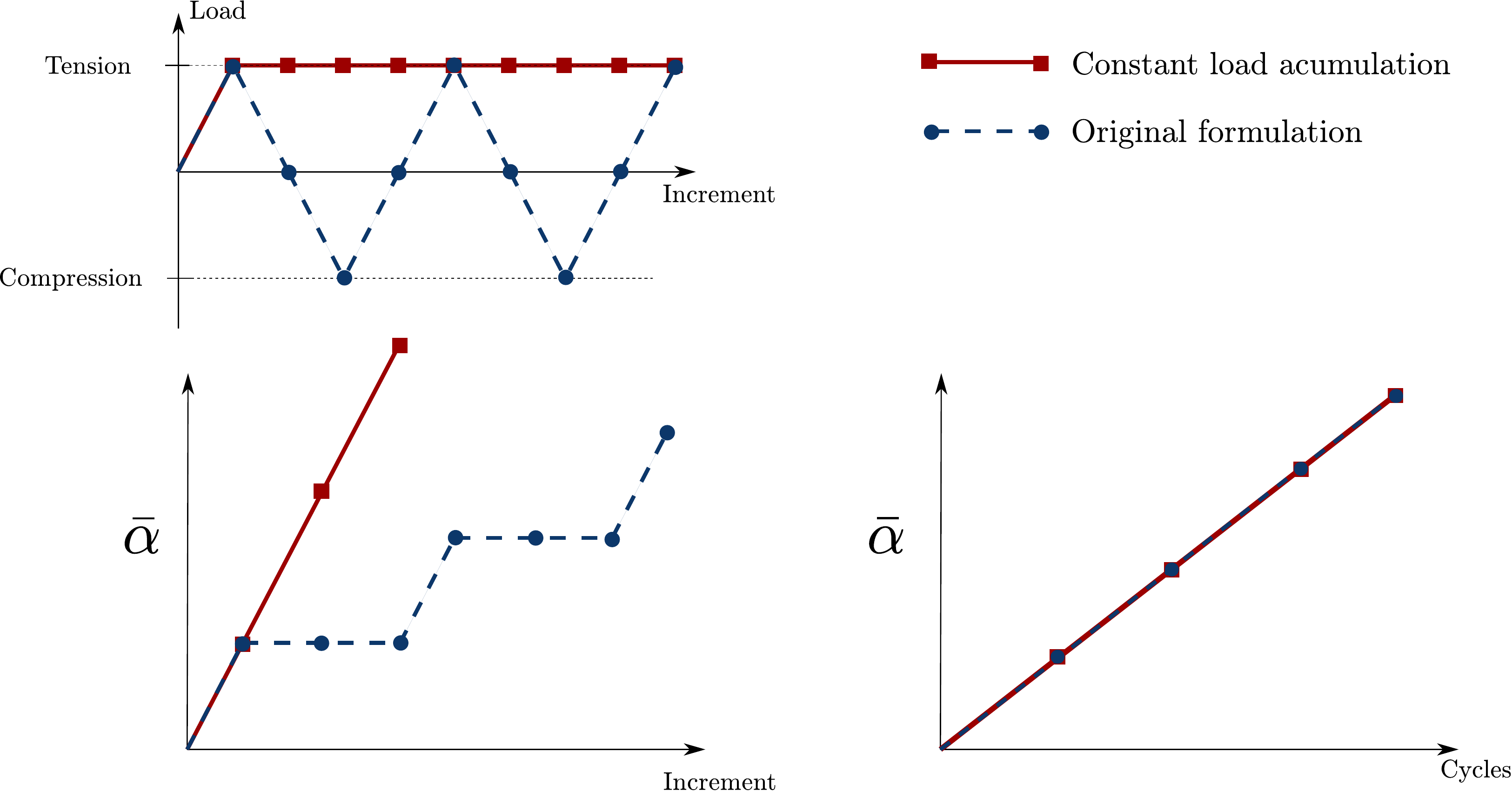}
    \caption{Illustration of loading pattern and fatigue accumulation for the original formulation on the left and the proposed constant load accumulation scheme of the right. As illustrated, the former requires at least four increments per load cycle, while the latter requires only one.}
    \label{fig:CLA}
\end{figure}

Another technique for accelerating cycle-by-cycle fatigue computations can be achieved by changing the way in which fatigue is accumulated. Here, henceforth referred to as the Constant Load Accumulation (CLA) acceleration strategy. In the current fatigue model, fatigue is accumulated by positive increments of active strain energy. The load ratio $R$ is here loosely defined in term of applied displacement $\Bar u$ as 
\begin{equation}
    R  = \dfrac{\Bar{u}_{min}}{\Bar{u}_{max}}.
\end{equation}
In the case where $R\geq 0$, the load cycle can be resolved using only two load increments per cycle (loading and unloading). However, only one of these increments actually contributes to fatigue. In the case where $R < 0$, four increments per cycle are required (tensile loading, unloading, compressive loading and unloading). If the compressive loading stage does not contribute significantly to fatigue, only the tensile loading increment is relevant, as illustrated by the dashed blue curves in Fig. \ref{fig:CLA}. Note that these considerations are only intended for high-cycle fatigue where material response is usually well within the linear regime and no effects such as compressive plasticity would be expected to occur to a significant degree. It should also be noted that there are loading conditions where multiple points in the load cycle can be relevant, such as non-proportional loading. \\
For problems which are sufficiently simple such that only one increment of the load cycle is significant, considerable acceleration of the computation can be achieved by changing the fatigue accumulation from Eq. \refe{eq:Faccumulation} to 
\begin{equation}\label{eq:CLA}
       \Bar{\alpha}_{n+1}=\Bar{\alpha}_n+\psi^+_{0,n+1},
\end{equation}
and combining this with the application of a constant load $\Bar{u}= \Bar{u}_{max}$ and the counting of one cycle per increment. The approach is illustrated by solid red lines in Fig. \ref{fig:CLA}.
We note for completeness that formally, to account for cases where $R>0$, the formulation should be amended to 
\begin{equation}
     \Bar{\alpha}_{n+1}=\Bar{\alpha}_n+\psi^+_{0,n+1}\left[1-R^2H(R)\right],
\end{equation}
although such cases will not be considered here. 

\section{Numerical experiments}
\label{sec:Results}

We shall now present the results of our numerical experiments, aiming at benchmarking the performance of the two novel acceleration strategies proposed here: the Modified Newton (MN) method presented in section \ref{sec:ModifiedNewton} and the constant load accumulation (CLA) scheme described in Section \ref{sec:CLA}. For all the case studies considered, the material parameters are chosen as Young's modulus $E=210$ GPa, Poisson's ratio $\nu=0.3$, and toughness $G_c = 2.7\,\tm{kJ}/\tm{m}^2$. First, the growth of fatigue cracks in a Single Edge Notched Tension (SENT) specimen (Section \ref{eq:SENT}) is investigated to compare acceleration schemes and quantify gains relative to the reference solution system. Then, more complex cracking patters are simulated by addressing the nucleation and growth of cracks in an asymmetric three point bending sample containing multiple holes (Section\ref{Sec:Asymmetric}). Here, one of the objectives is to compare the crack trajectories obtained with different strain energy decomposition approaches. Finally, in Section \ref{Sec:3Dcasestudy}, the robustness of the model and its ability to simulate fatigue cracking in large scale 3D problems is demonstrated. 

\subsection{Fatigue crack growth on a Single Edge Notched Tension (SENT) specimen}
\label{eq:SENT}

\begin{figure}[h]
    \centering
     \begin{subfigure}[b]{0.28\textwidth}
    \includegraphics[width=\linewidth]{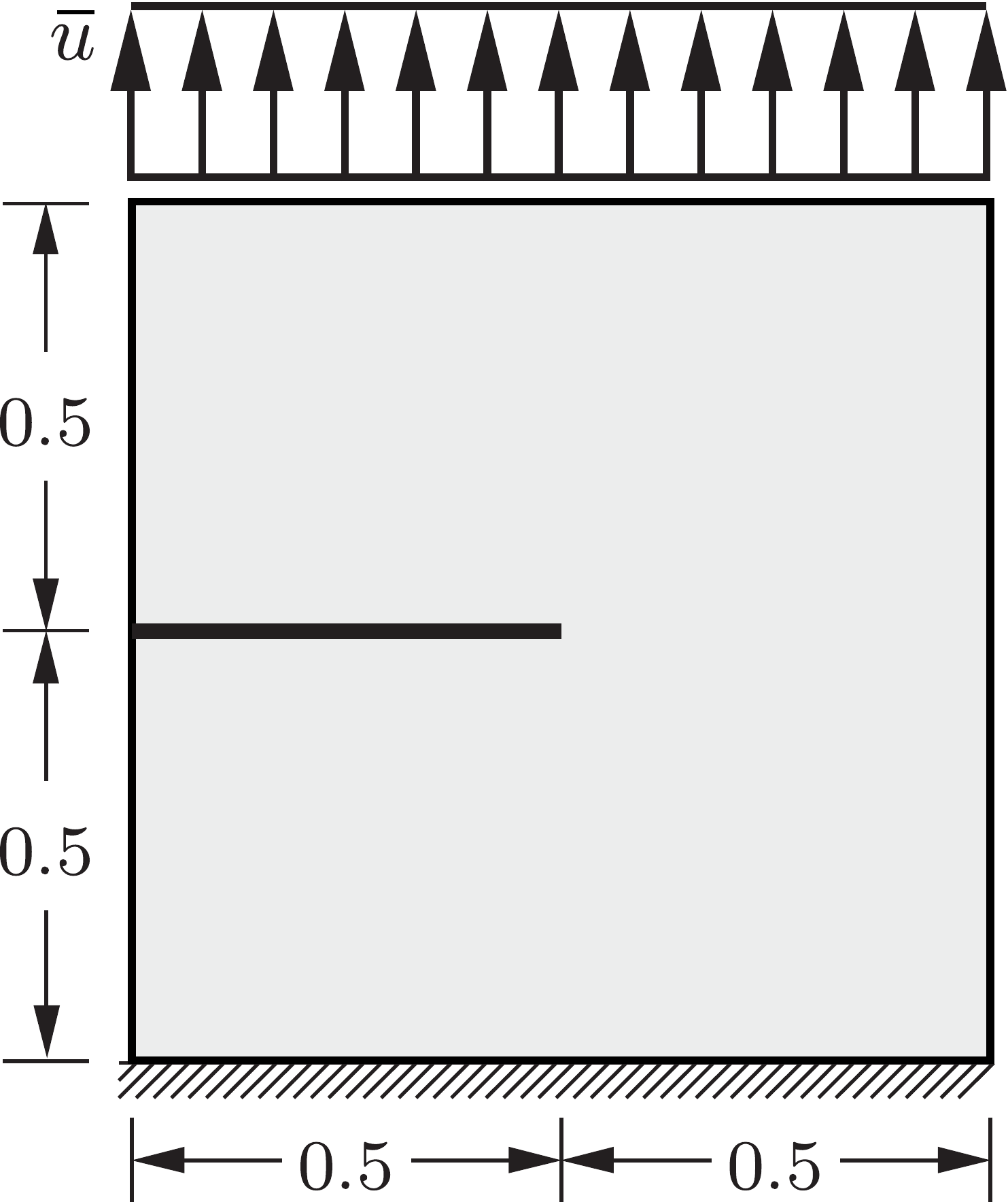}
    \caption{ }
    \label{fig:SENT_skecth}
    \end{subfigure}\hspace{1pt}
     \begin{subfigure}[b]{0.34\textwidth}
    \includegraphics[width=\linewidth]{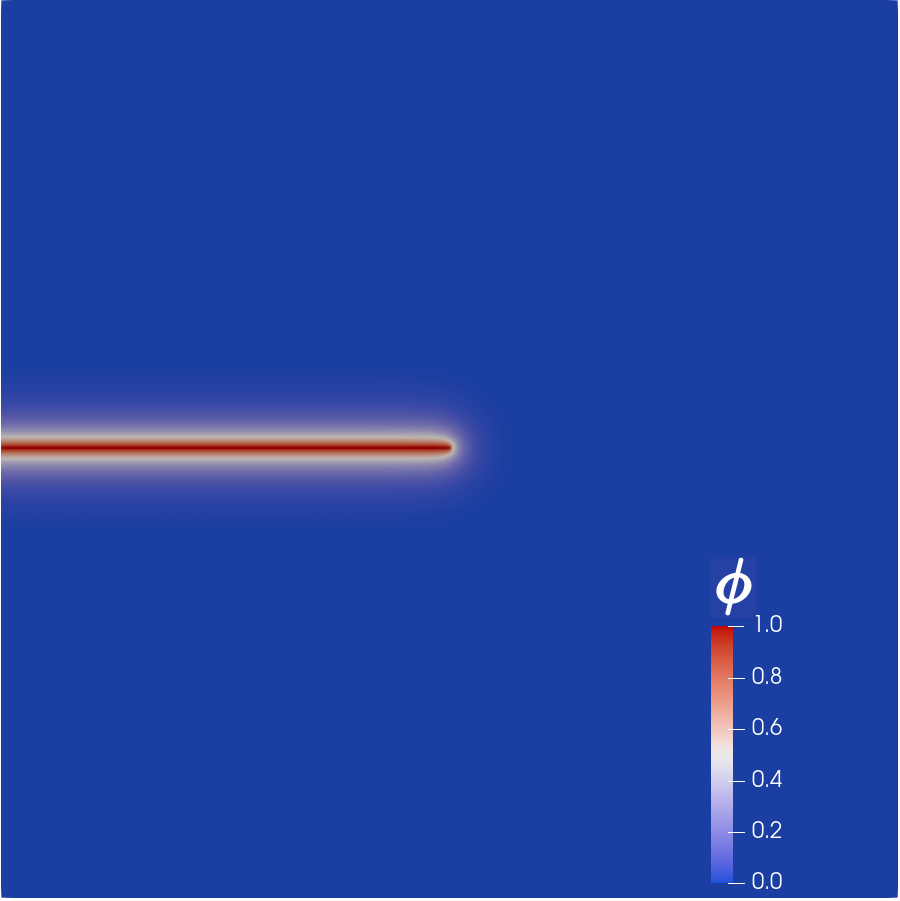}
    \caption{ }
    \label{fig:SENT_contour0}
    \end{subfigure}\hspace{1pt}
     \begin{subfigure}[b]{0.34\textwidth}
    \includegraphics[width=\linewidth]{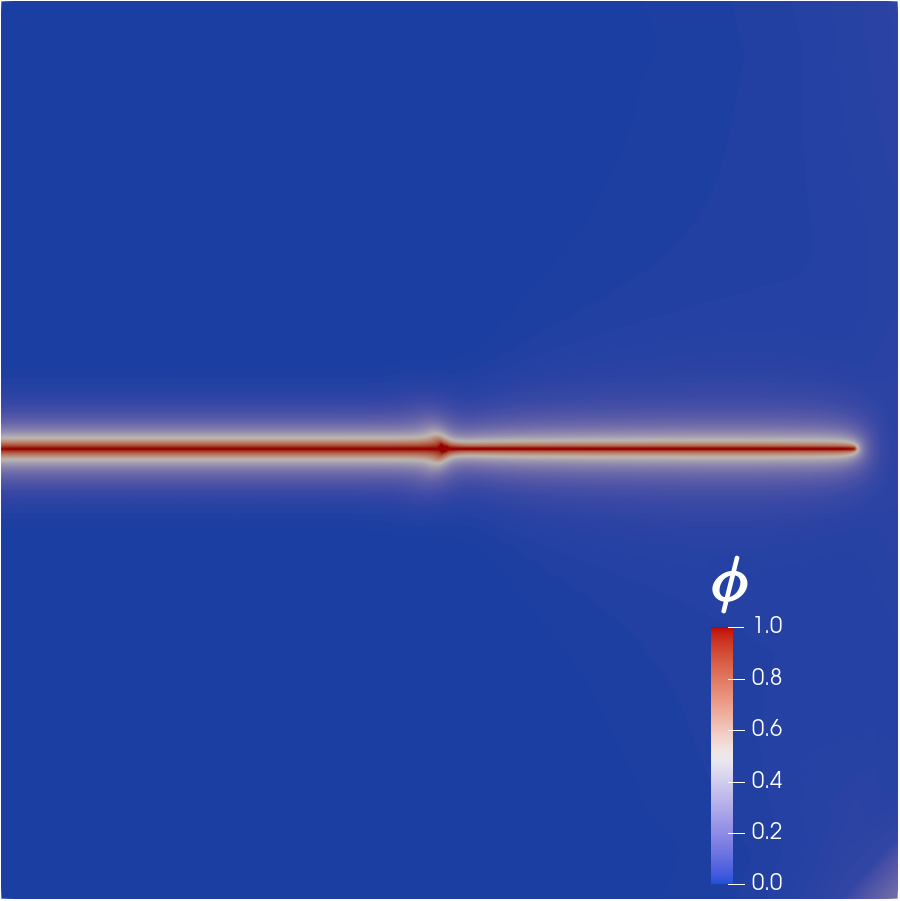}
    \caption{ }
    \label{fig:SENT_contour1}
    \end{subfigure}
    \caption{Fatigue crack growth in a Single Edge Notched Tension (SENT): (a) geometry (with dimensons in mm) and boundary conditions, (b) initial crack contour, and (c) final stage of the fatigue crack propagation.}
    \label{fig:SENT_
    introduction}
\end{figure}

The geometry and boundary conditions of the Single Edge Notched Tension (SENT) sample considered in the first case study are shown in Fig. \ref{fig:SENT_skecth}. The no-tension strain energy density decomposition given in Eq. \refe{eq:Freddi} is used and the initial crack is initialized as a Dirichlet condition on the phase field. The Dirichlet boundary condition is applied on two rows of elements so as to define a constant width for the initial crack.

The specimen is discretised using approximately 32,000 bilinear quadrilateral elements with a refined zone in the crack growth region. In this refined zone, the characteristic element length equals 0.003 mm, more than five times smaller than the phase field length scale, chosen here as $\ell = 0.016$ mm. The specimen is subjected to an alternating applied displacement $\Bar{u}$. The fatigue loading is repeated for 120.000 cycles with a maximum applied displacement $\Bar{u}_{max} = 0.0002$ mm. Calculations are obtained for four scenarios. First, results are obtained for the standard fatigue accumulation given in Eq. \refe{eq:Faccumulation} and the multipass staggered algorithm from Algorithm \ref{alg:Multipass}. These are considered to be the baseline conditions, not including any of the acceleration strategies proposed in this work. A second scenario constitutes the case where fatigue crack growth is simulated using the Modified Newton (MN) scheme presented in Section \ref{sec:ModifiedNewton}. Conversely, the third scenario employs only the Constant Load Acceleration (CLA) scheme described in Section \ref{sec:CLA}. Finally, a fourth scenario is considered where both MN and CLA approaches are used in tandem. The MN parameters are chosen as $n_i = 25$ and $n_c=100$ and the load ratio is always considered to be equal to $R=0$. All computations are performed with a single core of a CPU of the model Xeon E5-2650 v4. \\

\begin{figure}[h]
    \centering
    \includegraphics[width=0.8\textwidth]{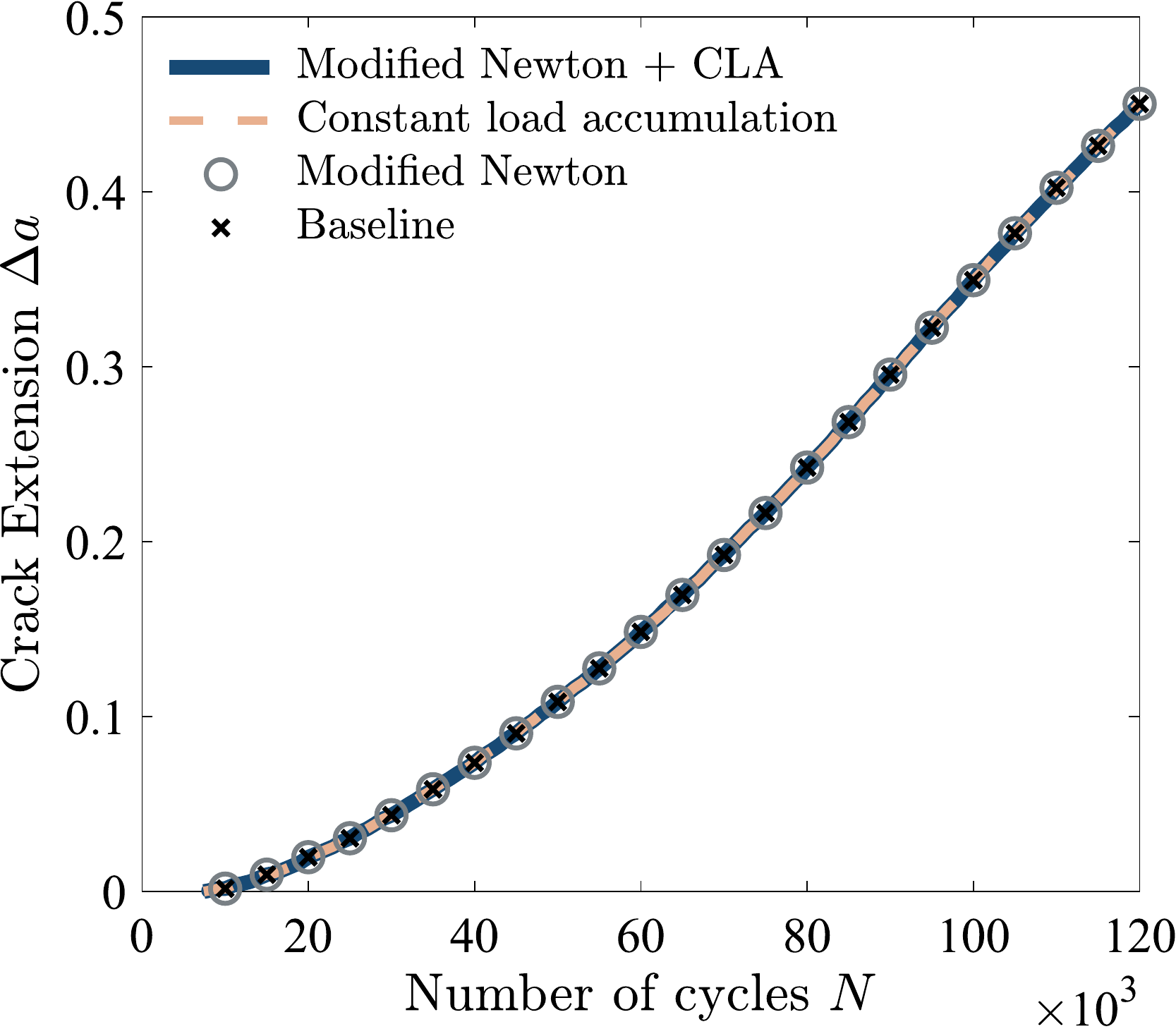}
    \caption{Predictions of crack extension $\Delta a$ versus number of cycles $N$ for the SENT case study. The figure shows results obtained with the reference (baseline) conditions, and the two acceleration scheme presented here (Modified Newton, NM; Constant load accumulation, CLA), independently and in tandem.}
    \label{fig:SENT_Extension}
\end{figure}

The finite element predictions of crack extension versus number of cycles are given in Fig. \ref{fig:SENT_Extension}. Here, crack extension is measured as the distance between the original crack tip and the furthest point with $\phi=0.95$. The results reveal that the acceleration schemes do not inherently introduce any deviation in crack extension when compared to the baseline. This is always the case for the modified Newton Method, and holds true for the CLA scheme when there is no compressive contribution to fatigue.\\

As shown in Table \ref{tab:SENT_performaance}, the computational performance of the different modelling strategies is measured by a number of factors. The first measure of performance is the actual computation time (in hours). However, one should note that although noise in this indicator has been minimized by the absence of parallel computing and the use of identical CPU types, individual measures of computation time should not be taken as an exact quantification of performance. A more objective measure is the second performance indicator considered, the total number of matrix factorizations, where one factorization here denotes a factorization of both the displacement and the damage subproblems. In addition, Table \ref{tab:SENT_performaance} also provides with the total number of iterations used on the phase field and displacement subproblems. The results reveal that while the use of a constant load accumulation acceleration strategy significantly reduces the computation time by reducing the necessary number of load increments, bringing a similar reduction in iterations and factorizations, the use of the Modified Newton approach presents a trade-off between a reduction in matrix factorizations and an increase in necessary iterations, especially on the displacement problem. However, with the choice of parameters for the modified Newton approach of $n_i=25$ and $n_c=100$, this strategy requires roughly 100 times less matrix factorizations than the baseline result in exchange for only 4 times more iterations on the displacement problem. Also, we emphasise that the iterations on the displacement problem are very cheap when using the MN approach as they only require rebuilding the residuals and trivially finding the solution with the existing factorized stiffness matrix. The results obtained show that both the Modified Newton (MN) and the constant load accumulation (CLA) accumulation strategies can provide substantial performance gains (independently or in tandem) without loss of accuracy. Moreover, these methods are compatible with existing cycle jump schemes such as those presented in \cite{Loew2020} and \cite{Seles2021}.

\begin{table}[h]
    \centering
    \begin{tabular}{ l| c  c  c |  c }
    Solutions strategy & MN + CLA & CLA & MN & Baseline\\\hline\hline
    Computation time [h]   & 13.0 & 32.7 & 40.1 & 96.5   \\ 
    Matrix factorizations   &  1191 & 120032 & 2379 & 240032 \\
    Total iterations $\phi$ & 120205 & 120032 & 240114 & 240032 \\
    Total iterations $\mathbf{u}$ & 276812 & 120032 & 973678 & 240032 \\\hline
    \end{tabular}
    \caption{Performance details for the SENT case study, comparing baseline results with the use of the Modified Newton (MN) approach, constant load accumulation (CLA) and a combination thereof.}
    \label{tab:SENT_performaance}
\end{table}

\subsubsection{Additional acceleration with multiple cycles per increment} \label{sec:CLA_N}

\begin{figure}[H]
    \centering
     \begin{subfigure}[b]{0.48\textwidth}
    \includegraphics[width=\linewidth]{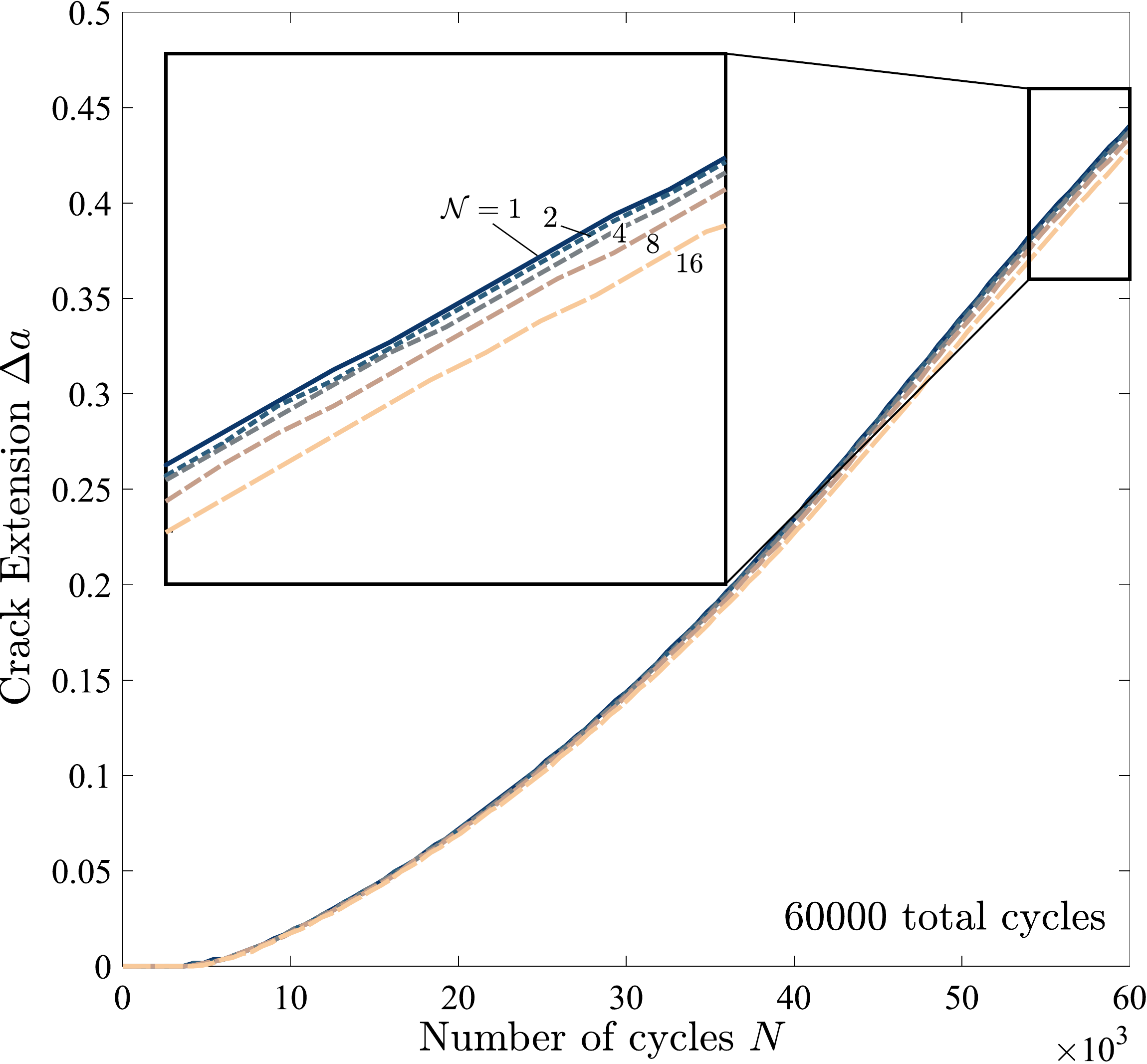}
    \caption{}
    \label{fig:SENT_60k}
    \end{subfigure}\hspace{4pt}
     \begin{subfigure}[b]{0.48\textwidth}
    \includegraphics[width=1.0\linewidth]{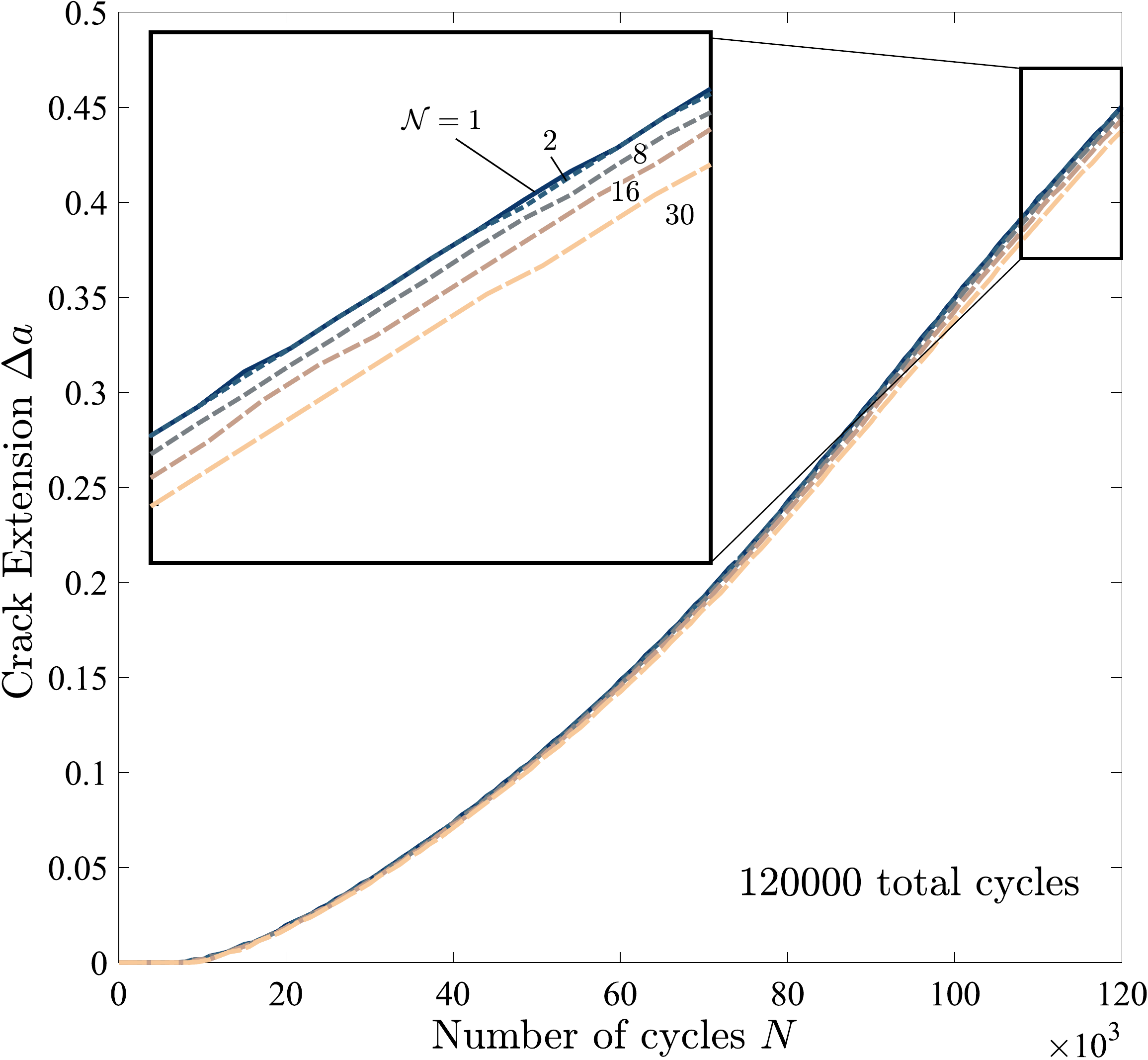}
    \caption{}
    \label{fig:SENT_120k}
    \end{subfigure}\
     \begin{subfigure}[b]{0.48\textwidth}
    \includegraphics[width=1.0\linewidth]{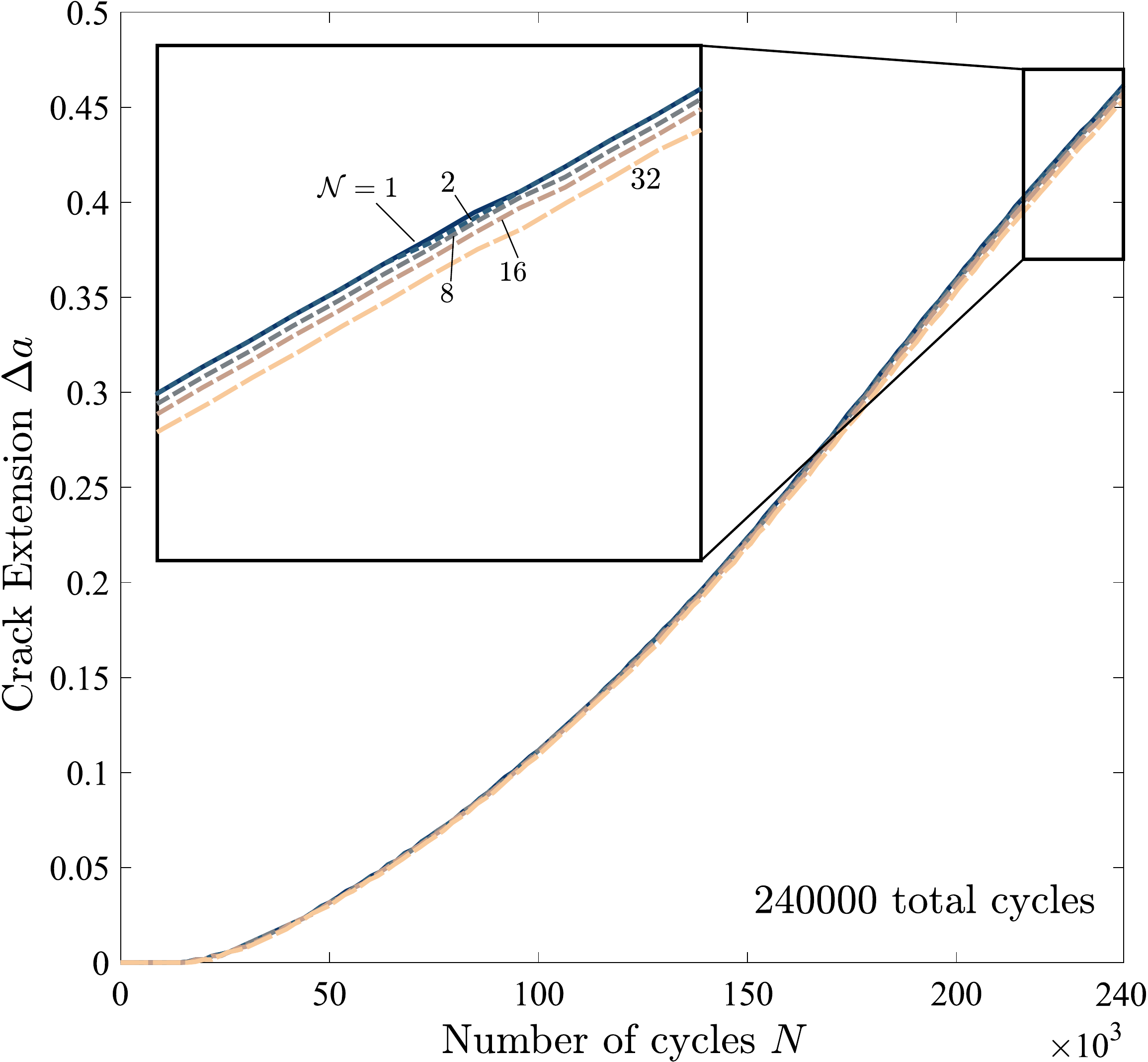}
    \caption{}
    \label{fig:SENT_240k}
    \end{subfigure}
    \caption{Predictions of crack extension $\Delta a$ versus number of cycles $N$ for the SENT case study considering selected values of $\mathcal{N}$ (number of cycles per increment) and the following characteristic number of cycles: (a) 60,000 cycles, (b) 120,000 cycles, and (c) 240,000 cycles.}
    \label{fig:SENT_acceleration}
\end{figure}

With the use of the constant load accumulation scheme, the choice of counting one cycle per increment is somewhat arbitrary as the update to the accumulated fatigue variable $\Bar{\alpha}$ from increment $n$ to increment $n+1$ can be readily modified to
\begin{equation}
     \Bar{\alpha}_{n+1}=\Bar{\alpha}_n+\mathcal{N}\psi_{0,n+1},
\end{equation}
where the number of cycles per increment $\mathcal{N}$ can be chosen smaller or greater than one. While a small degree of error is expected to arise as a result of the discrete sampling of the cycle history that will result from considering $\mathcal{N}>1$ (``cycle-jumping''), we here show that this error is negligible for simulations involving varying numbers of cycles to failure. To this end, calculations are conducted for the SENT geometry considering three values for the maximum applied displacement, namely $\Bar{u}_{max}=0.00016$ mm, $\Bar{u}_{max}=0.00020$ mm, and 
$\Bar{u}_{max}=0.00025$ mm, and corresponding characteristic number of cycles equal to $60,000$, $120,000$ and $240,000$, respectively. The number of cycles per increment $\mathcal{N}$ is varied in the range from $1$ to $32$ to investigate its influence on the accuracy of the solution. In all cases, this enhanced CLA approach is used in conjunction with the Modified Newton method. Results are provided for all three cases in Fig. \ref{fig:SENT_acceleration}. It can be observed that the accumulated error is small, with deviations in final crack extension from the reference $\mathcal{N}=1$ being in all cases below 3\%. Moreover, for a given value of $\mathcal{N}$, the deviation in the estimated crack extension at the end of the characteristic number of cycles decreases with increasing characteristic number of cycles. For high-cycle fatigue problems of engineering relevance, where the total number of cycles may be in the order of  millions, a high value of $\mathcal{N}$ can be used with insignificant error.  
Performance tables similar to Table \ref{tab:SENT_performaance} are provided in Appendix A. These show that, while computation times do not scale linearly with $\mathcal{N}$, they do monotonically decrease in the range investigated. Relative to the baseline cases in Table \ref{tab:SENT_performaance}, the corresponding computation with the Modified Newton method and the constant load accumulation technique with $\mathcal{N}=16$, is 32 times faster than the baseline case.

\subsection{Asymmetric three point bending}
\label{Sec:Asymmetric}

\begin{figure}[h]
    \centering
    \includegraphics[width=0.8\textwidth]{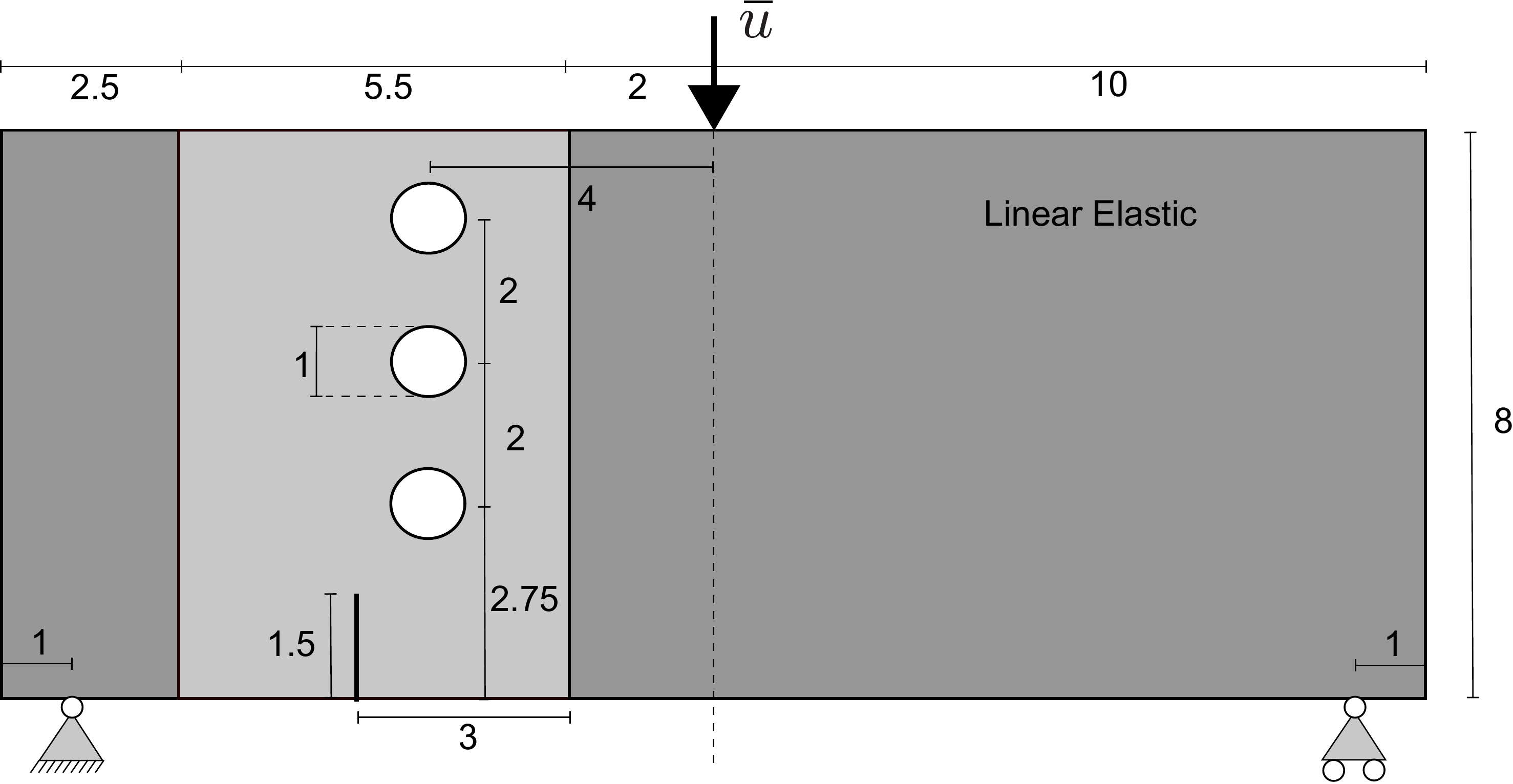}
    \caption{Sketch of the asymmetric three point bending problem, including dimensions (in mm) and boundary conditions. Only the light gray region is subject to fatigue damage.}
    \label{fig:TPB_sketch}
\end{figure}

The second case study aims at applying the fatigue acceleration schemes to a boundary value problem exhibiting more complex crack growth. As shown in Fig. \ref{fig:TPB_sketch}, a plane strain beam containing an array of holes is subjected to three point bending loading conditions. An initial crack is located asymmetric to the loading pins and the holes, inducing mixed-mode cracking. This paradigmatic boundary value problem has been previously investigated in the context of static loading (see, e.g., Refs. \cite{Molnar2017,Hirshikesh2019}). 

\begin{figure}[H]
    \centering
     \begin{subfigure}[b]{0.24\textwidth}
    \includegraphics[width=\linewidth]{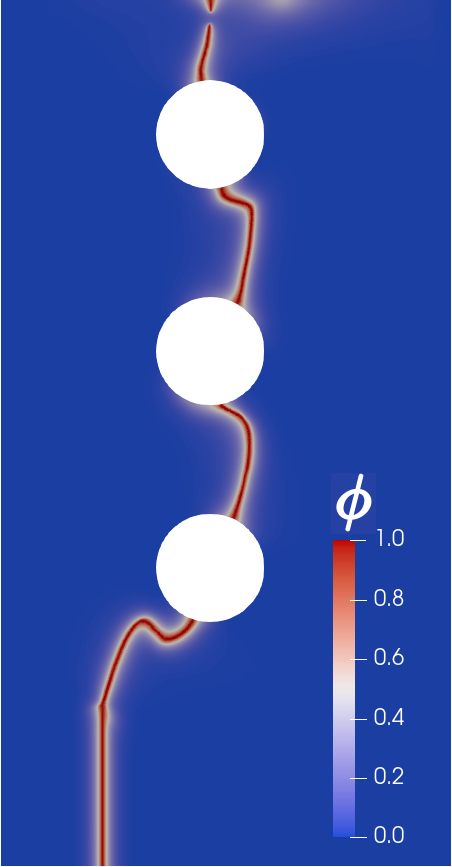}
    \caption{Isotropic}
    \label{fig:TPB_isotropic}
    \end{subfigure}\hspace{1pt}
     \begin{subfigure}[b]{0.24\textwidth}
    \includegraphics[width=\linewidth]{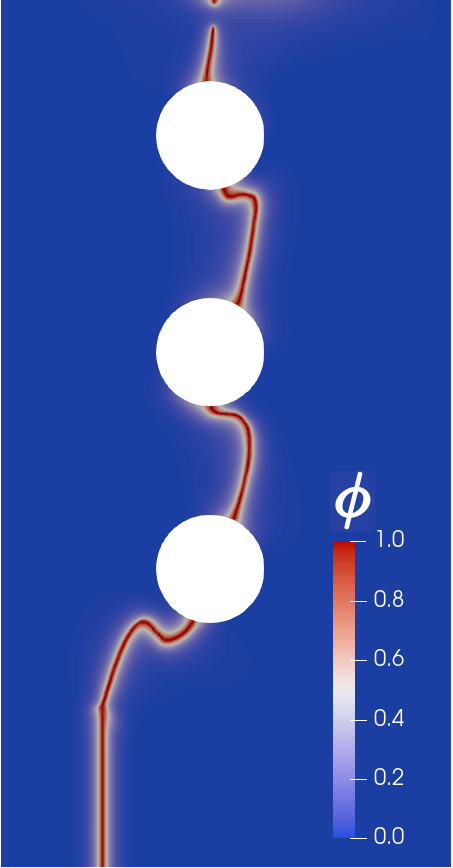}
    \caption{Vol/Dev}
    \label{fig:TPB_Amor}
    \end{subfigure}\hspace{1pt}
     \begin{subfigure}[b]{0.24\textwidth}
    \includegraphics[width=1.00\linewidth]{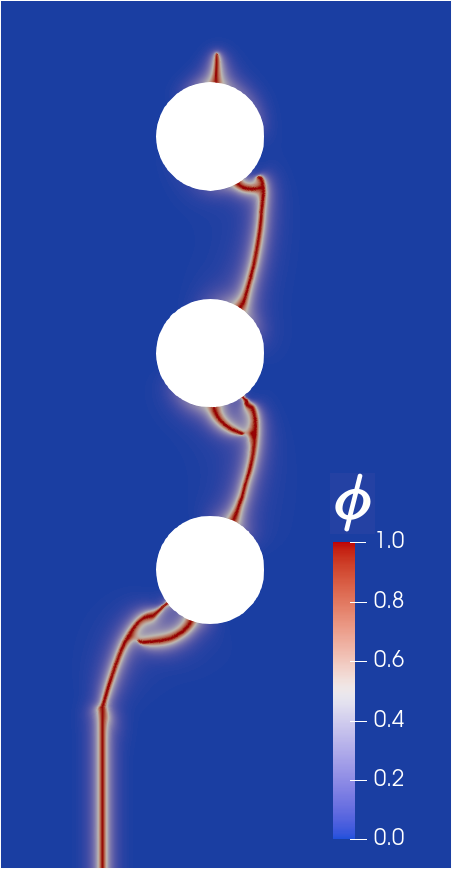}
    \caption{Spectral}
    \label{fig:TPB_Miehe}
    \end{subfigure}\hspace{1pt}
    \begin{subfigure}[b]{0.24\textwidth}
    \includegraphics[width=\linewidth]{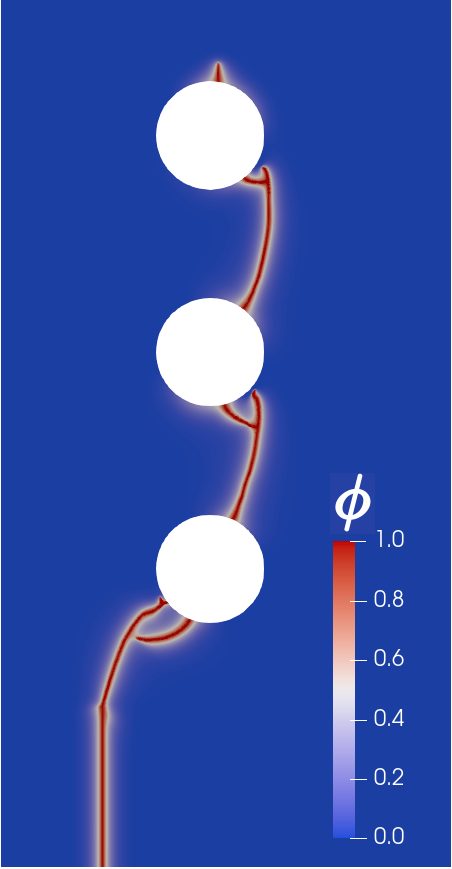}
    \caption{No-tension}
    \label{fig:TPB_Freddi}
    \end{subfigure}
    \caption{Asymmetric three point bending: Contours of the phase field crack growth after 90,000 fatigue cycles using a) no strain decomposition. b) The volumetric/deviatoric strain decomposition of \cite{Amor2009}. c) The spectral decomposition of \cite{Miehe2010a}. d) The no-tension split of \cite{Freddi2010}.}
    \label{fig:TPB_contours}
\end{figure}

The domain is discretised using roughly 128,000 linear quadrilateral elements, with a characteristic element size $h_e = 0.01$ mm, five times smaller than the phase field length scale. The applied displacement is $\Bar{u}=0.003$ mm, which is cycled for 90,000 load cycles. As is typical for three point bending experiments, the load ratio is assumed to be $R=0$. Both the modified Newton and the constant load accumulation scheme are exploited to capture the fatigue history in an efficient manner. The computation is carried out using different strain energy density decompositions to the differences in crack trajectory predictions. The obtained crack paths at the end of the analysis are shown in Fig. \ref{fig:TPB_contours}. 

We note that in all cases the crack path differs from those observed under monotonic loading; see, e.g., Refs. \cite{Molnar2017,Mandal2019}. This deviation is explained by the accumulation of fatigue near the holes which leads to the nucleation of new secondary cracks prior to the intersection of the primary crack with the holes. If an endurance limit were to be introduced in the phase field fatigue formulation, these secondary cracks could be eliminated and the monotonic loading crack path might be recovered. We also remark that the issues with nucleation of cracks from non-sharp defects highlighted by Strobl and Seelig \cite{Strobl2020}, which stems from the use of the history field approach for crack irreversibility, are not of significance for this phase field fatigue model. Performance measures for the four computations are provided in Table \ref{tab:TPB_performance}.

\begin{table}[H]
    \centering
    \begin{tabular}{ l| c  c  c c | c}
    Strain decomposition & Isotropic & Volumetric/deviatoric & Spectral &No-tension & No-tension* \\\hline\hline
    Computation time [hr]   & 44.9 & 44.4 & 60.3 & 75.5 & 399.0  \\ 
    Matrix factorizations   &  1249 & 1278 & 1341 & 1295 & 180130\\
    Total iterations $\phi$ & 91355 & 91400 & 91742 & 91624  & 180130\\
    Total iterations $\mathbf{u}$ & 314292 & 328948 & 400795 & 426645 & 180130 \\\hline
    \end{tabular}
    \caption{Performance details for the asymmetric three point bending case study as a function of the strain energy decomposition.
    *Without acceleration schemes. }
    \label{tab:TPB_performance}
\end{table}

The spectral and the no-tension split exhibit a more complex crack pattern and also require significantly more iterations on the displacement problem. However, in all cases, the proposed solution strategy offers a large reduction in the number of matrix factorizations required, in exchange for a modest increase in the number of iterations needed on the displacement problem. The results of this case study show that this performance improvement prevails even for complex crack growth studies. In the case of the No-tension split, which took the longest to compute with the modified Newton approach, it is still more than five times faster than when computed with a standard Newton method and without constant load accumulation. It can be expected that in the absence of these acceleration schemes, computation time is roughly independent of the strain decomposition as only one iteration per field per increment is required even in this most advanced case. Thus, the acceleration is roughly a factor of 9 for the isotropic and volumetric/deviatoric splits. As it was the case for the SENT specimen, a more optimal performance can most likely be achieved by differentiating how often the stiffness is updated for the two subproblems, as the damage subproblem can be updated far less frequently without paying the price of a significant number of additional iterations. 

\subsection{3D Beam under tension with a tilted edge crack} 
\label{Sec:3Dcasestudy}

\begin{figure}[h]
    \centering
    \includegraphics[width=1\textwidth]{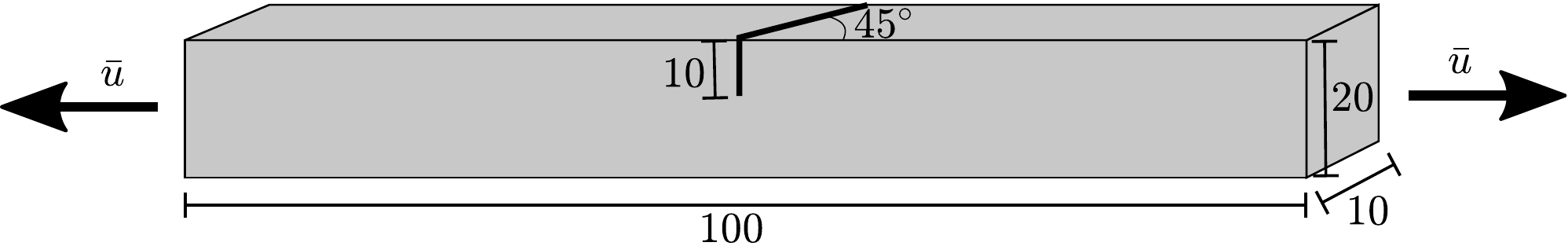}
    \caption{Sketch of the three-dimensional beam undergoing uniaxial tension and containing a tilted edge crack, including the sample dimensions (in mm) and boundary conditions.}
    \label{fig:3D_sketch}
\end{figure}

As a final benchmark, we consider the uniaxial tension of a three-dimensional beam with an edge crack. The induced complex crack behaviour, the crack is rotated $45^\circ$ relative to the beam cross section, as sketched in Fig. \ref{fig:3D_sketch}. 

\begin{figure}[H]
    \centering
     \begin{subfigure}[b]{0.32\textwidth}
    \includegraphics[width=0.9\linewidth]{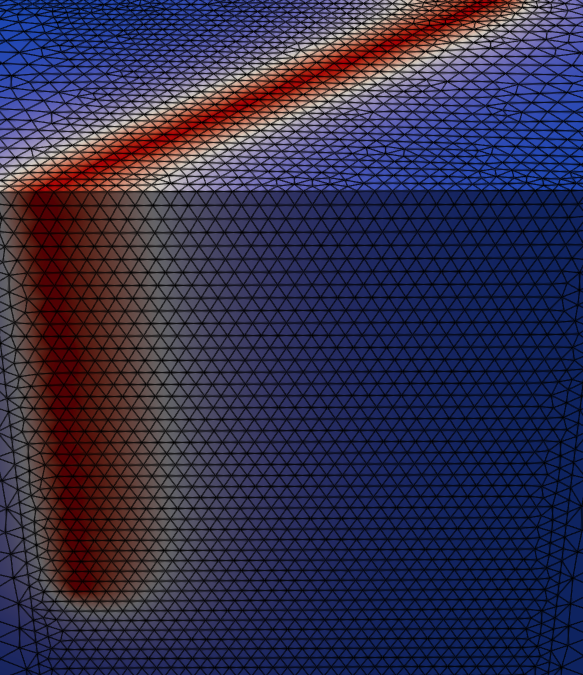}
    \caption{Mesh near the crack}
    \label{fig:3D_0}
    \end{subfigure}\hspace{1pt}
     \begin{subfigure}[b]{0.32\textwidth}
    \includegraphics[width=1\linewidth]{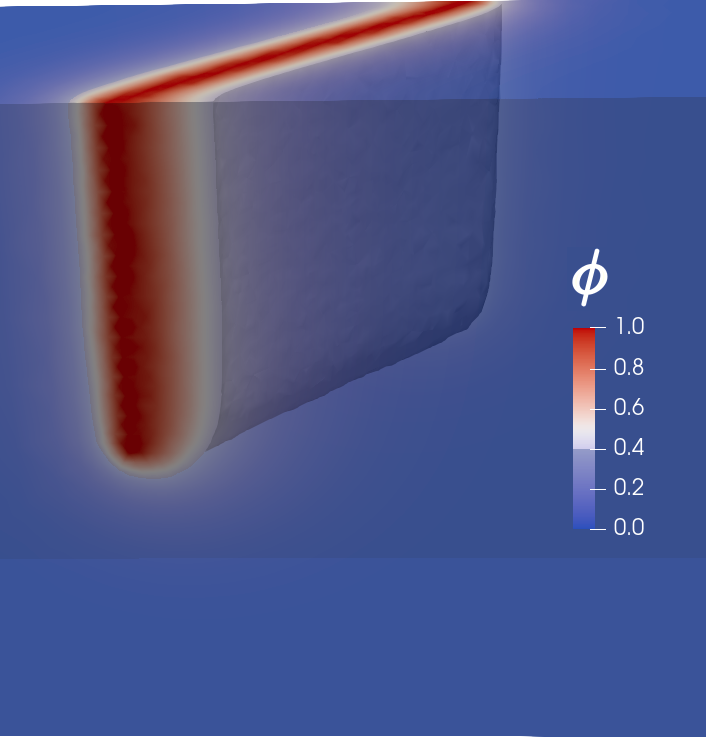}
    \caption{Initial state}
    \label{fig:3D_40k}
    \end{subfigure}\hspace{1pt}
     \begin{subfigure}[b]{0.32\textwidth}
    \includegraphics[width=0.965\linewidth]{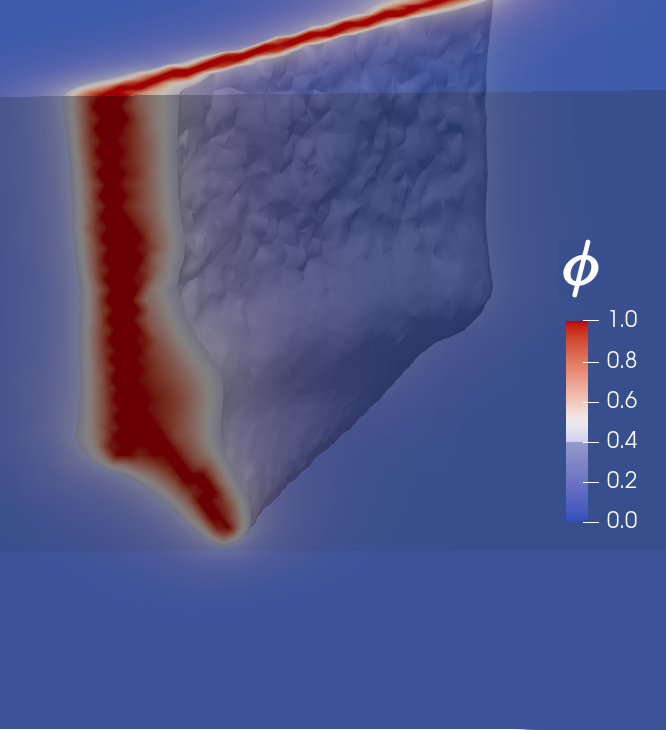}
    \caption{240000 cycles}
    \label{fig:3D_60k}
    \end{subfigure}\hspace{1pt}\\[0.5cm]
    
    \begin{subfigure}[b]{0.32\textwidth}
    \includegraphics[width=1.00\linewidth]{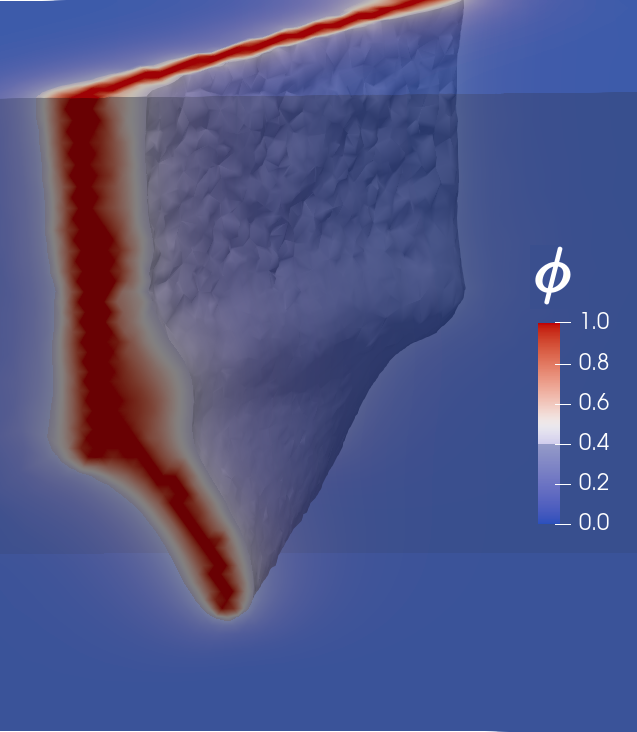}
    \caption{400000 cycles}
    \label{fig:3D_80k}
    \end{subfigure}
       \begin{subfigure}[b]{0.32\textwidth}
    \includegraphics[width=0.975\linewidth]{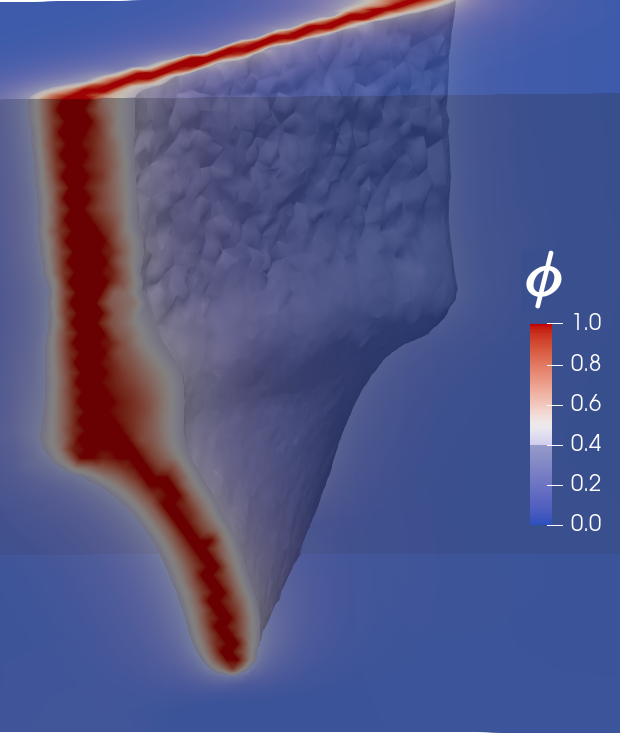}
    \caption{600000 cycles}
    \label{fig:3D_60k}
    \end{subfigure}\hspace{1pt}
    \begin{subfigure}[b]{0.32\textwidth}
    \includegraphics[width=\linewidth]{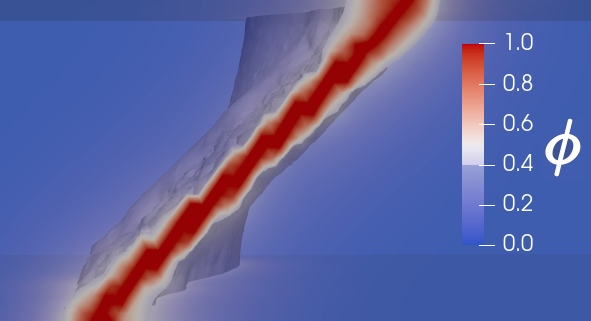}
    \caption{Top view - 600000 cycles}
    \label{fig:3D_80k}
    \end{subfigure}
    \caption{Contours of the 3D tension problem. The crack rotates to be perpendicular to the overall maximal principal stress.}
    \label{fig:3D_contours}
\end{figure}

The beam is subjected to cyclic tension by means of a displacement boundary condition applied on both ends. The load amplitude is $\Bar{u}=1$ mm and a total of 600 000 cycles are computed combining the Modified Newton approach with $n_i = 25$ and $n_c = 50$ with the constant load accumulation scheme with $\mathcal{N}=4$ cycles per increment (see Section \ref{sec:CLA_N}). The computational domain is meshed using approximately 196,000 linear tetrahedral elements, with a characteristic length near the crack $h_e = 0.35$ mm. The phase field length scale is here chosen to be equal to $\ell=1.2$ mm. The results obtained are shown in Fig. \ref{fig:3D_contours} in terms of the phase field contours, showing the crack growth pattern. 

The results obtained reveal the expected crack growth behaviour, with the crack rotating to position itself perpendicular to the overall maximum principal stress. The proposed methods accelerate the computation significantly, with a total of only 150,000 load increments needed to capture 600,000 cycles during which only 978 matrix factorizations are performed. In the absence of the Modified Newton approach and Constant Load Accumulation (CLA) acceleration strategies, a minimum of 1,200,000 increments would be required, with 1,200,000 matrix factorizations and 1,200,000 iterations on each subproblem. Combining the two acceleration schemes proposed here, the computation is achieved using only about 172,000 iterations for the displacement problem and 160,000 for the phase field problem. Furthermore, the majority of these iterations take significantly less time than in the baseline case, as only a few of them require matrix factorization. The number of matrix factorizations, which is the quantity expected to dominate the computation time for large problems, is reduced by a factor of over 400 in total. For a problem with a million cycles, a larger number of cycles per increment $\mathcal{N}$ can be employed for even greater computational improvement without loss of accuracy. Endowed with the acceleration strategies presented in this work, phase field fatigue is shown to be a technologically-relevant tool capable of delivering complex fatigue crack growth predictions in 3D over a hundred thousand cycles. 

\section{Concluding remarks}
\label{sec:Conclusions}

We have presented two compelling yet simple methods for accelerating phase field fatigue computations: (i) a Modified Newton (MN) approach, which is shown capable of drastically reducing the number of matrix factorizations necessary in the solution of a high cycle fatigue problem, and (ii) a constant load accumulation (CLA) approach that significantly reduces the number of load increments needed by considering only those relevant to the evolution of the fatigue variable. Three case studies are investigated to explore the performance benefits of these two acceleration strategies, individually and in tandem. Fatigue crack growth is predicted in 2D and 3D scenarios and compared with the baseline model. The results showed that computation times can be reduced by orders of magnitude when using MN and CLA and that these techniques remain robust even in the case of complex crack patterns and three-dimensional crack growth. The acceleration schemes presented enable predicting complex cracking patterns in 3D for over a hundred thousand cycles, endowing phase field fatigue models with the ability of delivering predictions for scales relevant to engineering practice. Moreover, the proposed methods are compatible other accelerations methods such as the cycle jump scheme presented in Ref. \cite{Loew2020}, which unlocks the potential for even greater performance benefits.

\section*{Acknowledgements}
The authors gratefully acknowledge financial support from the Danish Offshore Technology Centre (DHRTC) under the ”Structural Integrity and Lifetime Evaluation” programme as well as. Furthermore, this project would not have been possible without the expertise and resources of the DTU Computing Center \cite{DTU_DCC_resource}. Emilio Mart{\' i}nez-Pa{\~ n}eda additionally acknowledges financial support from UKRI’s Future Leaders Fellowship programme [grant MR/V024124/1]. A. Golahmar acknowledges financial support from Vattenfall Vindkraft A/S and Innovation Fund Denmark (grant 0153-00018B).

\section*{Appendix A: Performance data for the SENT specimen with multiple cycles per increment}

We here provide additional performance data for the computations addressed in Section \ref{sec:CLA_N}. Specifically, the performance data for the analysis with a characteristic number of cycles equal to 60,000 is given in Table \ref{tab:SENT_60k}, the data pertaining to the analysis for 120,000 cycles is given in Table \ref{tab:SENT_120k}, and the results for the 240,000 cycles case is provided in Table \ref{tab:SENT_240k}. Consistent with the main text, a matrix factorization denotes a factorization of the tangent stiffness matrix of both the damage and the displacement subproblems. The crack extension at the end of the total number of cycles for a given number of cycles per increment $\mathcal{N}$ is denoted $a_N$. Crack extension deviation is here measure relative to the $\mathcal{N}=1$ case such that the relative deviation $\Delta a_N$ is given by: 
\begin{equation}
    \Delta a_N = \dfrac{a_N - a_1}{a_1}.
\end{equation}
\begin{table}[H]
    \centering
    \begin{tabular}{ l| c  c  c c c}
    $\mathcal{N}$ & $1$ & $2$ & $4$ & $8$ & $16$ \\\hline\hline
    Computation time [hr]   & 8.7 & 6.8& 4.0 & 2.2 & 1.2  \\ 
    Matrix factorizations   &  606 & 333 & 298 & 294 & 290 \\
    Total iterations $\phi$ & 60314 & 30491 & 15600 & 8131 & 4528 \\
    Total iterations $\mathbf{u}$ & 234934 & 210754 & 126138 & 71018 & 39680 \\
    Crack extension deviation [\%]& - & 0.19 & -0.64 & -1.36 & -2.9 \\\hline
    \end{tabular}
    \caption{Performance details for the SENT specimen at a characteristic number of cycles of 60000. Computations utilize the proposed modified Newton approach and the constant load accumulation scheme with $N$ cycles per increment.}
    \label{tab:SENT_60k}
\end{table}

\begin{table}[H]
    \centering
    \begin{tabular}{ l| c  c  c c c}
    $\mathcal{N}$ & $1$ & $2$ & $8$ & $16$ & $30$ \\\hline\hline
    Computation time [hr]   & 13.0 & 9.3 & 5.3 & 3.0 & 1.3  \\ 
    Matrix factorizations   &  1191 & 609 & 293 & 297 & 293 \\
    Total iterations $\phi$ & 120205 & 60312 & 15634 & 8271 & 4778 \\
    Total iterations $\mathbf{u}$ & 276812 & 227920 & 127541 & 70568 & 40896 \\
    Crack extension deviation [\%] & - & 0.19 & -0.89 & -1.51 & -2.85\\\hline
    \end{tabular}
    \caption{Performance details for the SENT specimen at a characteristic number of cycles of 120,000. Computations utilize the proposed modified Newton approach and the constant load accumulation scheme with $N$ cycles per increment.}
    \label{tab:SENT_120k}
\end{table}

\begin{table}[H]
    \centering
    \begin{tabular}{ l| c  c  c c c}
    $\mathcal{N}$ & $1$ & $2$ & $8$ & $16$ & $32$ \\\hline\hline
    Computation time [hr]   & 20.6 & 12.4 & 7.0 & 4.0 & 2.2  \\ 
    Matrix factorizations   &  2378 & 1191 & 320 & 290 & 298 \\
    Total iterations $\phi$ & 240150 & 120227 & 30501 & 15614 & 8186 \\
    Total iterations $\mathbf{u}$ & 357149 & 266701 & 216611 & 127401 & 70764\\
    Crack extension deviation [\%] & - & 0.00 & -0.43 & -0.83 & -1.74\\\hline
    \end{tabular}
    \caption{Performance details for the SENT specimen at a characteristic number of cycles of 240000. Computations utilize the proposed modified Newton approach and the constant load accumulation scheme with $N$ cycles per increment.}
    \label{tab:SENT_240k}
\end{table}



\begin{thebibliography}{10}
\expandafter\ifx\csname url\endcsname\relax
  \def\url#1{\texttt{#1}}\fi
\expandafter\ifx\csname urlprefix\endcsname\relax\def\urlprefix{URL }\fi
\expandafter\ifx\csname href\endcsname\relax
  \def\href#1#2{#2} \def\path#1{#1}\fi

\bibitem{Bourdin2000}
B.~Bourdin, G.~A. Francfort, J.~J. Marigo, {Numerical experiments in revisited
  brittle fracture}, Journal of the Mechanics and Physics of Solids 48~(4)
  (2000) 1--23.

\bibitem{Francfort1998}
G.~A. Francfort, J.~J. Marigo, {Revisiting Brittle Fracture As an Energy
  Minimization Problem}, Journal of the Mechanics and Physics of Solids 46~(8)
  (1998) 1319--1342.

\bibitem{Borden2012}
M.~J. Borden, C.~V. Verhoosel, M.~A. Scott, T.~J. Hughes, C.~M. Landis, {A
  phase-field description of dynamic brittle fracture}, Computer Methods in
  Applied Mechanics and Engineering 217-220 (2012) 77--95.

\bibitem{Kristensen2020b}
P.~K. Kristensen, C.~F. Niordson, E.~Mart{\'{i}}nez-Pa{\~{n}}eda, {Applications
  of phase field fracture in modelling hydrogen assisted failures}, Theoretical
  and Applied Fracture Mechanics 110 (2020).

\bibitem{Hirshikesh2019}
Hirshikesh, S.~Natarajan, R.~K. Annabattula, E.~Mart{\'{i}}nez-Pa{\~{n}}eda,
  {Phase field modelling of crack propagation in functionally graded
  materials}, Composites Part B: Engineering 169 (2019) 239--248.

\bibitem{Tanne2018}
E.~Tann{\'{e}}, T.~Li, B.~Bourdin, J.~J. Marigo, C.~Maurini, {Crack nucleation
  in variational phase-field models of brittle fracture}, Journal of the
  Mechanics and Physics of Solids 110 (2018) 80--99.

\bibitem{Kristensen2021}
P.~K. Kristensen, C.~F. Niordson, E.~Mart{\'{i}}nez-Pa{\~{n}}eda, {An
  assessment of phase field fracture: Crack initiation and growth},
  Philosophical Transactions of the Royal Society A: Mathematical, Physical and
  Engineering Sciences 379~(2203) (2021).

\bibitem{Navidtehrani2022}
Y.~Navidtehrani, C.~Beteg{\'{o}}n, E.~Mart{\'{i}}nez-Pa{\~{n}}eda, {A general
  framework for decomposing the phase field fracture driving force,
  particularised to a Drucker–Prager failure surface}, Theoretical and
  Applied Fracture Mechanics 121 (2022) 103555.

\bibitem{DeLorenzis2022}
L.~D. Lorenzis, C.~Maurini, Nucleation under multi-axial loading in variational
  phase-field models of brittle fracture, International Journal of Fracture 237
  (2022) 61--81.

\bibitem{Wu2017}
J.-Y. Wu, {A unified phase-field theory for the mechanics of damage and
  quasi-brittle failure}, Journal of the Mechanics and Physics of Solids 103
  (2017) 72--99.

\bibitem{Feng2022}
Y.~Feng, J.~Li, {Phase-Field cohesive fracture theory: A unified framework for
  dissipative systems based on variational inequality of virtual works},
  Journal of the Mechanics and Physics of Solids 159 (2022) 104737.

\bibitem{Guillen-Hernandez2020}
T.~Guill{\'{e}}n-Hern{\'{a}}ndez, A.~Quintana-Corominas, I.~G. Garc{\'{i}}a,
  J.~Reinoso, M.~Paggi, A.~Tur{\'{o}}n, {In-situ strength effects in long fibre
  reinforced composites: A micro-mechanical analysis using the phase field
  approach of fracture}, Theoretical and Applied Fracture Mechanics 108 (2020)
  102621.

\bibitem{Tan2021}
W.~Tan, E.~Mart{\'{i}}nez-Pa{\~{n}}eda, {Phase field predictions of microscopic
  fracture and R-curve behaviour of fibre-reinforced composites}, Composites
  Science and Technology 202 (2021) 108539.

\bibitem{Aldakheel2018}
F.~Aldakheel, P.~Wriggers, C.~Miehe, {A modified Gurson-type plasticity model
  at finite strains: formulation, numerical analysis and phase-field coupling},
  Computational Mechanics 62~(4) (2018) 815--833.

\bibitem{Alessi2018}
R.~Alessi, M.~Ambati, T.~Gerasimov, S.~Vidoli, L.~D. Lorenzis, {Comparison of
  Phase-Field Models of Fracture Coupled with Plasticity}, Computational
  Methods in Applied Sciences 46 (2018) 1--21.

\bibitem{Bourdin2014}
B.~Bourdin, J.-J. Marigo, C.~Maurini, P.~Sicsic, {Morphogenesis and Propagation
  of Complex Cracks Induced by Thermal Shocks}, Physical Review Letters 112~(1)
  (2014) 14301.

\bibitem{Ye2022}
J.~Y. Ye, L.~W. Zhang, {Damage evolution of polymer-matrix multiphase
  composites under coupled moisture effects}, Computer Methods in Applied
  Mechanics and Engineering 388 (2022) 114213.

\bibitem{Martinez-Paneda2018}
E.~Mart{\'{i}}nez-Pa{\~{n}}eda, A.~Golahmar, C.~F. Niordson, {A phase field
  formulation for hydrogen assisted cracking}, Computer Methods in Applied
  Mechanics and Engineering 342 (2018) 742--761.

\bibitem{Duda2018}
F.~P. Duda, A.~Ciarbonetti, S.~Toro, A.~E. Huespe, {A phase-field model for
  solute-assisted brittle fracture in elastic-plastic solids}, International
  Journal of Plasticity 102~(November 2017) (2018) 16--40.

\bibitem{Anand2019}
L.~Anand, Y.~Mao, B.~Talamini, {On modeling fracture of ferritic steels due to
  hydrogen embrittlement}, Journal of the Mechanics and Physics of Solids 122
  (2019) 280--314.

\bibitem{Kristensen2020}
P.~K. Kristensen, C.~F. Niordson, E.~Mart{\'{i}}nez-Pa{\~{n}}eda, {A phase
  field model for elastic-gradient-plastic solids undergoing hydrogen
  embrittlement}, Journal of the Mechanics and Physics of Solids 143 (2020)
  104093.

\bibitem{Klinsmann2016}
M.~Klinsmann, D.~Rosato, M.~Kamlah, R.~M. McMeeking, {Modeling crack growth
  during Li insertion in storage particles using a fracture phase field
  approach}, Journal of the Mechanics and Physics of Solids 92 (2016) 313--344.

\bibitem{Boyce2022}
A.~M. Boyce, E.~Martínez-Pañeda, A.~Wade, Y.~S. Zhang, J.~J. Bailey, T.~M.
  Heenan, D.~J. Brett, P.~R. Shearing, Cracking predictions of lithium-ion
  battery electrodes by x-ray computed tomography and modelling, Journal of
  Power Sources 526 (2022) 231119.

\bibitem{Ai2022}
W.~Ai, B.~Wu, E.~Martínez-Pañeda, A coupled phase field formulation for
  modelling fatigue cracking in lithium-ion battery electrode particles,
  Journal of Power Sources 544 (2022) 231805.

\bibitem{Alessi2023}
R.~Alessi, J.~Ulloa, Endowing griffith’s fracture theory with the ability to
  describe fatigue cracks, Engineering Fracture Mechanics (2023) 109048.

\bibitem{Carrara2020}
P.~Carrara, M.~Ambati, R.~Alessi, L.~{De Lorenzis}, {A framework to model the
  fatigue behavior of brittle materials based on a variational phase-field
  approach}, Computer Methods in Applied Mechanics and Engineering 361 (2020)
  112731.

\bibitem{Loew2020}
P.~J. Loew, L.~H. Poh, B.~Peters, L.~A. Beex, {Accelerating fatigue simulations
  of a phase-field damage model for rubber}, Computer Methods in Applied
  Mechanics and Engineering 370 (2020) 113247.

\bibitem{Golahmar2022}
A.~Golahmar, P.~K. Kristensen, C.~F. Niordson, E.~Mart{\'{i}}nez-Pa{\~{n}}eda,
  {A phase field model for hydrogen-assisted fatigue}, International Journal of
  Fatigue 154 (2022) 106521.

\bibitem{Simoes2021}
M.~Simoes, E.~Mart{\'{i}}nez-Pa{\~{n}}eda, {Phase field modelling of fracture
  and fatigue in Shape Memory Alloys}, Computer Methods in Applied Mechanics
  and Engineering 373 (2021) 113504.

\bibitem{Simoes2022}
M.~Simoes, C.~Braithwaite, A.~Makaya, E.~Martínez-Pañeda, Modelling fatigue
  crack growth in shape memory alloys, Fatigue \& Fracture of Engineering
  Materials \& Structures 45~(4) (2022) 1243--1257.

\bibitem{Seiler2019}
M.~Seiler, T.~Linse, P.~Hantschke, M.~Kästner, An efficient phase-field model
  for fatigue fracture in ductile materials, Engineering Fracture Mechanics 224
  (2020) 106807.

\bibitem{Mesgarnejad2019}
A.~Mesgarnejad, A.~Imanian, A.~Karma, Phase-field models for fatigue crack
  growth, Theoretical and Applied Fracture Mechanics 103 (2019) 102282.

\bibitem{Song2022}
J.~Song, L.~G. Zhao, H.~Qi, S.~Li, D.~Shi, J.~Huang, Y.~Su, K.~Zhang, Coupling
  of phase field and viscoplasticity for modelling cyclic softening and crack
  growth under fatigue, European Journal of Mechanics - A/Solids 92 (2022)
  104472.

\bibitem{Lo2019}
Y.~S. Lo, M.~J. Borden, K.~Ravi-Chandar, C.~M. Landis, {A phase-field model for
  fatigue crack growth}, Journal of the Mechanics and Physics of Solids 132
  (2019) 103684.

\bibitem{Gerasimov2016}
T.~Gerasimov, L.~{De Lorenzis}, {A line search assisted monolithic approach for
  phase-field computing of brittle fracture}, Computer Methods in Applied
  Mechanics and Engineering 312 (2016) 276--303.

\bibitem{Heister2015}
T.~Heister, M.~F. Wheeler, T.~Wick, {A primal-dual active set method and
  predictor-corrector mesh adaptivity for computing fracture propagation using
  a phase-field approach}, Computer Methods in Applied Mechanics and
  Engineering 290 (2015) 466--495.

\bibitem{Klinsmann2015}
M.~Klinsmann, D.~Rosato, M.~Kamlah, R.~M. McMeeking, {An assessment of the
  phase field formulation for crack growth}, Computer Methods in Applied
  Mechanics and Engineering 294 (2015) 313--330.

\bibitem{Freddi2022}
F.~Freddi, L.~Mingazzi, Mesh refinement procedures for the phase field approach
  to brittle fracture, Computer Methods in Applied Mechanics and Engineering
  388 (2022) 114214.

\bibitem{Freddi2023}
F.~Freddi, L.~Mingazzi, Adaptive mesh refinement for the phase field method: A
  fenics implementation, Applications in Engineering Science 14 (2023) 100127.

\bibitem{Olesch2021}
D.~Olesch, C.~Kuhn, A.~Schl{\"{u}}ter, R.~M{\"{u}}ller, {Adaptive numerical
  integration of exponential finite elements for a phase field fracture model},
  Computational Mechanics 67~(3) (2021) 811--821.

\bibitem{Sargado2021}
J.~M. Sargado, E.~Keilegavlen, I.~Berre, J.~M. Nordbotten, {A combined finite
  element–finite volume framework for phase-field fracture}, Computer Methods
  in Applied Mechanics and Engineering 373 (2021) 113474.

\bibitem{Ambati2016}
M.~Ambati, R.~Kruse, L.~{De Lorenzis}, {A phase-field model for ductile
  fracture at finite strains and its experimental verification}, Computational
  Mechanics 57~(1) (2016) 149--167.

\bibitem{Seles2019}
K.~Sele{\v{s}}, T.~Lesi{\v{c}}ar, Z.~Tonkovi{\'{c}}, J.~Sori{\'{c}}, {A
  residual control staggered solution scheme for the phase-field modeling of
  brittle fracture}, Engineering Fracture Mechanics 205 (2019) 370--386.

\bibitem{Kristensen2020a}
P.~K. Kristensen, E.~Mart{\'{i}}nez-Pa{\~{n}}eda, {Phase field fracture
  modelling using quasi-Newton methods and a new adaptive step scheme},
  Theoretical and Applied Fracture Mechanics 107 (2020) 102446.

\bibitem{Wu2019}
J.-y.~Y. Wu, Y.~Huang, V.~Phu, V.~P. Nguyen, {On the BFGS monolithic algorithm
  for the unified phase field damage theory}, Computer Methods in Applied
  Mechanics and Engineering 360 (2020) 112704.

\bibitem{Lampron2021}
O.~Lampron, D.~Therriault, M.~L{\'{e}}vesque, {An efficient and robust
  monolithic approach to phase-field brittle fracture using a modified Newton
  method}, Computer Methods in Applied Mechanics and Engineering 306 (2021).

\bibitem{Borjesson2022}
E.~B{\"{o}}rjesson, J.~J. Remmers, M.~Fagerstr{\"{o}}m, {A generalised
  path-following solver for robust analysis of material failure}, Computational
  Mechanics (2022).

\bibitem{JodlBauer2020}
D.~Jodlbauer, U.~Langer, T.~Wick, {Matrix-free multigrid solvers for
  phase-field fracture problems}, Computer Methods in Applied Mechanics and
  Engineering 372 (2020) 113431.

\bibitem{Cojocaru2006}
D.~Cojocaru, A.~M. Karlsson, {A simple numerical method of cycle jumps for
  cyclically loaded structures}, International Journal of Fatigue 28~(12)
  (2006) 1677--1689.

\bibitem{Miehe2010}
C.~Miehe, M.~Hofacker, F.~Welschinger, {A phase field model for
  rate-independent crack propagation: Robust algorithmic implementation based
  on operator splits}, Computer Methods in Applied Mechanics and Engineering
  199~(45-48) (2010) 2765--2778.

\bibitem{Ambati2014}
M.~Ambati, T.~Gerasimov, L.~{De Lorenzis}, {A review on phase-field models of
  brittle fracture and a new fast hybrid formulation}, Computational Mechanics
  55~(2) (2014) 383--405.

\bibitem{Seles2021}
K.~Sele{\v{s}}, F.~Aldakheel, Z.~Tonkovi{\'{c}}, J.~Sori{\'{c}}, P.~Wriggers,
  {A general phase-field model for fatigue failure in brittle and ductile
  solids}, Computational Mechanics 67~(5) (2021) 1431--1452.

\bibitem{Amor2009}
H.~Amor, J.~J. Marigo, C.~Maurini, {Regularized formulation of the variational
  brittle fracture with unilateral contact: Numerical experiments}, Journal of
  the Mechanics and Physics of Solids 57~(8) (2009) 1209--1229.

\bibitem{Miehe2010a}
C.~Miehe, F.~Welschinger, M.~Hofacker, {Thermodynamically consistent
  phase-field models of fracture: Variational principles and multi-field FE
  implementations}, International Journal for Numerical Methods in Engineering
  83 (2010) 1273--1311.

\bibitem{Golahmar2023}
A.~Golahmar, C.~F. Niordson, E.~Martínez-Pañeda, A phase field model for
  high-cycle fatigue: Total-life analysis, International Journal of Fatigue 170
  (2023) 107558.

\bibitem{Freddi2010}
F.~Freddi, G.~Royer-Carfagni, {Regularized variational theories of fracture: A
  unified approach}, Journal of the Mechanics and Physics of Solids 58~(8)
  (2010) 1154--1174.

\bibitem{Carlsson_Ferrite_jl_2021}
K.~Carlsson, F.~Ekre, Contributors,
  \href{https://github.com/Ferrite-FEM/Ferrite.jl}{{Ferrite.jl}} (3 2021).
\newline\urlprefix\url{https://github.com/Ferrite-FEM/Ferrite.jl}

\bibitem{Linse2017}
T.~Linse, P.~Hennig, M.~K{\"{a}}stner, R.~de~Borst, {A convergence study of
  phase-field models for brittle fracture}, Engineering Fracture Mechanics 184
  (2017) 307--318.

\bibitem{Strobl2020}
M.~Strobl, T.~Seelig, {Phase field modeling of Hertzian indentation fracture},
  Journal of the Mechanics and Physics of Solids 143 (2020).

\bibitem{Molnar2017}
G.~Moln{\'{a}}r, A.~Gravouil, {2D and 3D Abaqus implementation of a robust
  staggered phase-field solution for modeling brittle fracture}, Finite
  Elements in Analysis and Design 130 (2017) 27--38.

\bibitem{Mandal2019}
T.~K. Mandal, V.~P. Nguyen, J.~Y. Wu, {Length scale and mesh bias sensitivity
  of phase-field models for brittle and cohesive fracture}, Engineering
  Fracture Mechanics 217 (2019) 106532.

\bibitem{DTU_DCC_resource}
{DTU Computing Center}, {DTU Computing Center resources} (2021).

\end{thebibliography}
\end{document}